\documentclass[useAMS,usenatbib,usegraphicx]{mn2e}
\input epsf
\newcommand{\etal}{{\it et~al.}}

\newcommand{\mstar}{$M_\odot$}
\newcommand{\microm}{$\mu$m}

\newcommand{\MSX}{{\it MSX}}
\newcommand{\uchii}{{UC H{\scriptsize II}}}
\newcommand{\hii}{{H{\scriptsize II}}}

\newcommand{\mjycount}{$\mbox{mJy cnt}^{-1}$}

\title[SCUBA observations of cold cores \& the dust emissivity index $\beta$]{Millimetre continuum observations of southern massive star formation regions.\\
{\bf II.  SCUBA observations of cold cores and the dust grain emissivity index ($\beta$)}}

\author[T. Hill \etal~]{T. Hill$^1$\thanks{Email:thill@phys.unsw.edu.au}, M. A. Thompson$^2$, M. G. Burton$^1$, A. J. Walsh$^1$, V. Minier$^3$, \and M. R. Cunningham$^1$, and D. Pierce-Price$^4$\\
\\
$^{1}$ School of Physics, University of New South Wales, Sydney 2052, NSW, Australia.\\
$^{2}$ Centre for Astrophysics Research, Science \& Technology Research Institute, University of Hertfordshire, College lane,\\ Hatfield, Herts, AL10 9AB, UK\\
$^{3}$ Service d'Astrophysique, DAPNIA/DSM/CEA CE de Saclay, 91, 191 Gif-sur-Yvette, France.\\
$^{4}$ Joint Astronomy Centre, 660 N. A\'{}ohoku Pl, University Park, Hilo Hawaii 96720, USA\\
}

\begin{document}

\date{Accepted/ Received}

\pagerange{\pageref{firstpage}--\pageref{lastpage}} \pubyear{2006}

\maketitle

\label{firstpage}

\begin{abstract}
%72 targeted are mm-only, 1 is maser,  3 mr, 2 radio

   We report the results of a submillimetre continuum emission survey targeted toward 78 star formation regions, 72 of which are devoid of methanol maser and \uchii\, regions, identified in the SEST/SIMBA millimetre continuum survey of \citet{hill05}. At least 45 per cent of the latter sources, dubbed `mm-only', detected in this survey are also devoid of mid infrared \MSX\, emission. The 450 and 850\microm\, continuum emission was mapped using the Submillimetre Common User Bolometer Array (SCUBA) instrument on the James Clerk Maxwell Telescope (JCMT). Emission is detected toward 97 per cent of the 78 sources targeted as well as towards 28 other SIMBA sources lying in the SCUBA fields.

  In total, we have identified 212 cores in this submillimetre survey, including 106 previously known from the SIMBA survey. Of the remaining 106 sources, 53 result from resolving a SIMBA source into multiple submillimetre components, whilst the other 53 sources are  submillimetre cores, not seen in SIMBA. Additionally, we have identified two further mm-only sources in the SIMBA images. Of the total 405 sources identified in the SIMBA survey, 255 are only seen at millimetre wavelengths.

   We concatenate the results from four (sub)millimetre continuum surveys of massive star formation \citep[][as well as this work]{walsh03, hill05, thompson06}, together with the Galactic Plane map of \citet{p-p00} in order to determine the dust grain emissivity index $\beta$ for each of the sources in the SIMBA source list. We examine the value of $\beta$ with respect to temperature, as well as for the source classes identified in the SIMBA survey, for variation of this index. Our results indicate that $\beta$ is typically 2, which is consistent with previous determinations in the literature, but for a considerably larger sample than previous work.

\end{abstract}
\section{Introduction}

   Massive star formation tends to occur associated or coincident with methanol maser sites and/or \uchii\, regions (\citealp{batrla87, caswell95, walsh97}; \citealp*{minier01}; \citealp{beuther02, faundez04}; \citealp*{williams04}; \citealp{thompson06}, and references within each). A recent SIMBA millimetre continuum emission study toward these typical star formation identifiers by \citet[][hereafter Paper I]{hill05}, revealed evidence of bright, massive cores which were devoid of methanol maser and/or radio continuum emission. These sources, detected solely from their millimetre continuum emission, were dubbed `mm-only' cores. Preliminary analysis revealed these mm-only sources to be smaller and less massive, on average, than those sources displaying evidence of methanol maser and/or radio continuum emission, although they have similar total mass ranges. It is not yet clear which of the mm-only cores will form massive stars, although even the least massive of the mm-only cores in the sample have sufficient mass to support stars in excess of 10\mstar\, forming. It is therefore almost certain that at least some of these mm-only cores will produce massive clusters.

   In this paper we present the JCMT/SCUBA submillimetre continuum observations of mm-only cores detected in the SEST/SIMBA survey (Paper I). The primary aim of this submillimetre survey was to obtain data to complement that of SIMBA, in order to examine the nature of these mm-only cores, and to determine if they are associated with cold, deeply embedded emission. The SIMBA and SCUBA data combined will also allow examination of the dust grain emissivity index $\beta$ as explored in the following paragraphs.

   The combination of the 1.2\,mm, 850\microm\, and 450\microm\, data places important constraints on the grain properties of the source, in terms of the opacity and the types of grains that comprise the central star-forming complex. If the dust properties are constrained, then, through spectral energy distribution (SED) fitting, estimates of the dust temperature, bolometric luminosity and the dust mass can be obtained \citep{lis00}. Constraining the dust properties also allows inference of the star formation efficiency \citep*{hoare91}. In the absence of data at the peak of the SED curve ($\sim$100\microm), the 450\microm\, data is particularly important in constraining SED fits.

   Accurate dust temperatures (T$_d$) are difficult to estimate \citep{dunne00}. Dust radiates as a greybody - a modified Planck function with an extra emissivity term (Q $\propto \nu^\beta$), where $\beta$ is the dust grain emissivity exponent. The dust emissivity exponent is determined from the ratio of the submillimetre and millimetre observations. For a known value of $\beta$, these ratios are a sensitive function of temperature \citep{lis00}. (Sub)millimetre observations provide the most direct means for estimating the mass of the circumstellar or interstellar dust, since at these wavelengths, even the most massive clouds are optically thin \citep{emerson88}. Since protostellar cores get warmer as they evolve, this flux ratio also provides a diagnostic of the evolutionary stage of the core \citep{lis00}.

   Knowledge of the emissivity index improves estimates of the dust opacity which in turn bears heavily on the determination of dust mass from from measurement of the emission intensities. To obtain the mass, $\beta$ must be known, or assumptions about this exponent and the dust temperature need to be made. The dust grain emissivity index indicates the type of grain in the star forming core, as well as constraining the slope of spectral energy distribution (SED) fits.

   Typical values of $\beta$ range between 1 and 2, with Kramers-Kronig theorem constraining the lower limit at 1 \citep{jackson62, emerson88}. There is however, a wealth of observations as well as laboratory measurements for higher values of $\beta$ than 2 \citep*[c.f.][]{goldsmith97}, which suggests that the upper limit of this exponent is dependent on the environment, which in turn dictates the type of grains that can form.

   There are few large scale observational surveys of $\beta$ as per this one, instead many prefer to assume a value of $\beta$ in calculating the dust mass or SED fitting when $\beta$ is unknown \citep{dunne00, james02, kramer03, bianchi03}. Generally a value of 2 is assumed, indicative of graphitic grains \citep*{mennella95}. A review of work involving the dust grain emissivity exponent $\beta$ is included in the paper of \citet{goldsmith97}, as well as in Section \ref{chap:scuba:discuss:beta} here.

   In this paper, we present the observations and analysis of the submillimetre (450 \& 850\microm) SCUBA continuum survey obtained on the JCMT for mm-only sources reported in Paper I. The dust grain emissivity index $\beta$ is determined using the millimetre (1.2\,mm) data obtained from the SIMBA survey and the submillimetre data reported here, as well as existing SCUBA data \citep{p-p00, walsh03, thompson06} for the methanol maser and \uchii\, sources reported in Paper I. We examine the value of $\beta$ with respect to temperature and source class type, and present the results of this analysis here (Section \ref{chap:scuba:discussion}). The results from the analysis of $\beta$ will also constrain the dust properties (as described above) for spectral energy distribution analysis, to be presented in a forthcoming paper (Paper III). The primary aim of this paper is to present the results of the SCUBA submillimetre continuum survey toward the mm-only cores of Paper I, and to examine the dust grain emissivity exponent for a much larger sample of sources than has previously been considered.

\section[]{Observations and Data Reduction}
\subsection{The Sample}

   In Paper I we reported observations of a 1.2\,mm SIMBA continuum emission survey conducted on the SEST, toward star forming complexes harbouring a methanol maser and/or an \uchii\, region. The SIMBA source list was predominantly drawn from the existing SCUBA surveys of \citet{walsh03} and \citet{thompson06}, which were themselves originally targeted toward methanol masers and \uchii\, regions respectively.

   As a result of the large map sizes (240 $\times$ 480 arcsec$^2$) offered by the SEST, the SIMBA survey revealed numerous other cores quite clearly differentiated and offset from the targeted maser and radio sources. The SIMBA survey (Paper I) revealed 404 sources, of which $\sim$\,63 per cent were previously unknown. These `mm-only' cores are devoid of methanol maser and radio continuum sources, and at least 45 per cent are also without mid infrared \MSX\, emission. 

   The aim of this SCUBA survey was to obtain the complementary submillimetre continuum data for the mm-only cores detected in Paper I.

   Due to declination limits and time constraints, it was not possible to obtain complementary submillimetre SCUBA data for all of the 255 mm-only cores in the SIMBA sample of Paper I.

\subsection{Submillimetre Observations (450 \& 850\microm)}

   The submillimetre continuum observations were undertaken on the 15\,m James Clerk Maxwell Telescope (JCMT), using the Submillimetre Common User Bolometer Array (SCUBA) during two separate observing periods in 2003 August/September as well as in 2005 May.

   SCUBA \citep{holland99} is a submillimetre continuum array receiver with two hexagonal bolometer arrays, covering a field of view of 2.3 arcmin in diameter. The `long-wave' array which operates at 850\microm, is comprised of 37 pixels, whilst the `short-wave' array has 91 pixels and operates at 450\microm.  A dichroic beamsplitter allows these two arrays to be used simultaneously, and diffraction-limited resolution of each of the bolometers results in a resolution of 8 arscec at 450\microm, and 15 arcsec at 850\microm. 

   Observations were taken in jigglemap mode  \citep[see][]{holland99}. The hexagonal arrangement of the bolometers means that the sky is instantaneously undersampled. In jigglemap mode, the secondary mirror is moved in a 64-point jiggle pattern with 3 arcsec sampling, which essentially fills the gaps between the bolometers and is thus required to fully sample each of the arrays. Using the SIMBA maps from Paper I, we were, in most instances, able to set appropriate chop throws to avoid chopping onto nearby emission. Chop throws were typically 140 arcsec in azimuth.

   The initial SCUBA allocation of 10 hours was queue scheduled, and consequently spanned a period of 5 nights over the period 2003 August-September. During this period 24 regions were mapped simultaneously at 450 and 850\microm. Skydips were performed every 1-2 hours in order to correct for the atmospheric opacity. Opacities fluctuated on a nightly basis, with values typical of SCUBA grade three weather ($\tau_{225\, \rm GHz}$\,=\,0.08\,-\,0.12) for this period. Maps of Uranus were taken for flux calibration purposes.  Opacities as well as the calibration factor  for each of the nights of this period are listed in Table \ref{chap:scuba:calfactors}. Typical map integration times were 10 minutes on source. The average 1-$\sigma$ rms noise in the maps for this period is 3.7\,Jy at 450\microm\, and 0.2\,Jy at 850\microm.

   The main SCUBA observations were conducted during three second-half nights in 2005 May (5th-7th inclusive). During this period, observations spanned a total of 16 hours, allowing 54 regions to be mapped simultaneously at 450 and 850\microm. In order to accurately monitor the sky conditions, skydips were performed on an hourly basis, as were pointings. Sky opacities for this period are typical of SCUBA grade two weather ($\tau_{225\, \rm GHz}$\,=0.05\,-\,0.08). Maps of 16293-2422 were used to calibrate the data from the first two nights and CRL 2688 on the final night. Typical map integration times for this period were 10 minutes per source, resulting in an average residual noise for this period of 0.5\,Jy at 450\microm\, and 0.1\,Jy at 850\microm.

  Opacities for each night of observation over these two periods are listed in Table \ref{chap:scuba:calfactors} as well as the Flux Conversion Factor (FCF) determined from the images and the error beam correction (Beam Corr.).

\begin{table*} 
  \caption{Summary of the opacities, flux calibration factors (FCF), and the correction for the contribution of the error beam (Beam Corr.) for each night of the SCUBA observations. Note that a $\star$ denotes observations for which a suitable calibrator was not observed, and consequently the data can not be calibrated.\label{chap:scuba:calfactors} }
  \begin{center}
    \footnotesize
    \begin{tabular}{|lcccccc|}
      \hline
      &                   &             &  \multicolumn{2} {c} {450\microm} & \multicolumn{2} {c} {850\microm} \\         
      Date     & No. of   & Opacities   & FCF          & Beam   &  FCF        & Beam \\
      dd mm yy & Sources  &  $\tau_{225\, \rm GHz}$~~~~ &\mjycount~    & Corr.  &  \mjycount~ & Corr. \\
               & Obs &             & $beam^{-1}$       &        & $beam^{-1}$ &    \\
      \hline
      02 Aug 03 & 2      & 0.09-0.13   &     287.9   &1.50     &   221.5   & 1.18 \\
      04 Aug 03 & 2      & 0.08        &     253.7   &1.82     &   248.5   & 1.23 \\
      24 Aug 03 & 8     & 0.09        &     192.8   &1.48     &   243.0   & 1.13 \\
      17 Sep 03 & 4     & 0.08        &     290.1   &1.51     &   244.3   & 1.13 \\
      20 Sep 03 & 8     & 0.08        &     $\star$  &$\star$  &   318.9   & 1.13 \\
      %& & & \\
      05 May 05 & 24     & 0.06-0.07   &    443.0    &  1.38    &  249.6   &  1.18 \\
      06 May 05 & 22     & 0.06        &    287.9    &  1.51    &  227.7   &  1.45 \\
      07 May 05 & 8      & 0.05-0.07   &    317.2    &  1.12    &  256.0   &    1.07 \\
      \hline
    \end{tabular}
  \end{center}
\end{table*}

\subsection{Data Reduction and Analysis}

   The data were reduced and analysed using the standard SCUBA User Reduction Facility (SURF) package \citep{jenness98}, in conjunction with the Starlink image analysis packages KAPPA and GAIA. The data reduction procedure is as described in the SURF users manual \citep{jenness00}. In short, the data are subject to nod compensation, flatfielding, extinction correction, flagging, sky noise removal, bolometer weighting and despiking, prior to map creation. Noisy bolometers, which are evident in Figure \ref{chap:scuba:scubasample} as `holes' in the maps, were removed during this procedure.

   Flux calibration was performed according to the night that the data were taken. The 2003 data were calibrated using Uranus, whilst the 2005 data were calibrated using the secondary calibrators 16293-2422 and CRL\,2688, as the primary calibrator for this period, Mars, was resolved. The Starlink program FLUXES was used to obtain the expected flux for Uranus during the observation period. The flux of the secondary calibrator 16293-2422 was assumed to be 62.7 and 15.1 Jy/beam at 450 and 850\microm\,  respectively,  whilst CRL\,2688 was assumed to have a flux of 22.0 and 5.9\,Jy/beam at 450 and 850\microm\,  respectively\footnote{Fluxes were taken from the JCMT SCUBA documentation pages http://www.jach.hawaii.edu/JCMT/continuum/calibrat\-ion/sens/secondary\_2004.html.}.

  The flux calibration procedure was similar for both periods of observation for both the 450 and 850\microm\, data. Flux conversion factors  (FCFs) were determined from the ratio of the expected flux to the measured peak of the calibrator. Each map was then calibrated in Jy/beam by applying the appropriate FCF (see Table \ref{chap:scuba:calfactors}). The calibrated map was then converted to FITS file format. The resultant flux of the calibrator after application of the FCF was then compared to the known flux of the calibrator in order to determine the error beam correction, which is also listed in Table \ref{chap:scuba:calfactors}.

   The flux density of each source was determined using the Starlink package GAIA by applying an aperture around the source and at various points in the image considered to be the background. The flux inside each of the apertures was then measured and the background subtracted. Contour levels of 10 per cent  of the peak flux were overlaid on the sources in order to maintain source size consistency.

\subsection{Millimetre Observations (1.2\,mm)}

   The millimetre continuum observations were undertaken on the 15\,m Swedish ESO Submillimetre Telescope (SEST), using the SEST IMaging Bolometer Array (SIMBA) during three observing periods spanning 2001 October to 2002 October. SIMBA is a 37-channel hexagonal bolometer array with a central operating frequency of 250\,GHz (1.2\,mm), and a half power beam width of 24 arcsec for a single element with separation on the sky of 44 arcsec. Observations were taken using a fixed secondary mirror in a fast mapping mode, with typical integration times of 15\,min on source. The observational, calibration and data reduction procedures are as described in Paper I.

\section[]{Results}

\subsection{SCUBA}\label{chap:scuba:results:scuba}

   This SCUBA survey at 450 and 850\microm\, targeted a total of 78 sources from the SIMBA millimetre continuum survey (Paper I) over the two observing periods in 2003 and 2005. 72 of the sources targeted are mm-only sources devoid of methanol maser and radio continuum emission, whilst the other 6 display evidence of methanol maser and/or radio continuum emission. The images of these regions, as well as the flux density, for each of the sources detected in this survey are reported here. A summary of the results from this survey is given in Table \ref{chap:scuba:scubafindings}.

   In total, 212 sources are detected in this SCUBA survey, of which 106 were known from the SIMBA survey (where 78 were targeted, and the other 28 sources fell within the fields of the SCUBA maps but were not directly targeted here). The remaining 106 sources reported in Table \ref{chap:scuba:scubafluxes} are submillimetre cores of two distinct categories. The first category of 53 sources (50 per cent) arises as a result of the different resolutions offered by the SIMBA and SCUBA instruments, and are consequently the product of resolving known SIMBA cores into multiple components. This is discussed further in Section \ref{chap:scuba:combining}.

\begin{table*}
\caption{Summary of findings from the SCUBA survey.  For further explanation refer to the text (Section \ref{chap:scuba:results:scuba}). \label{chap:scuba:scubafindings}}
\begin{center}
\footnotesize
\begin{tabular}{|lcl|}
\hline
Total Number & &  Description\\
\hline
78 SIMBA targeted &=& 72 mm-only sources + 6 with methanol maser sites and/or \uchii\, regions.\\
&&\\
212 SCUBA observed & =& 106 -$>$78 targeted + 28 others in SIMBA survey within the maps, but not directly targeted.\\
                   & + &106 -$>$ 53 resolve known SIMBA sources into multiple components + 29 no SIMBA emission +\\
         & & 24 not directly associated with SIMBA but found within extended emission of a SIMBA core.\\
\hline
\end{tabular}
\end{center}
\end{table*}

   The second category of submillimetre source (53) are devoid of a known SIMBA millimetre continuum source. 24 of these sources fall within the extended emission of a nearby SIMBA source, with the majority located at the edge of this emission. The remaining 29 submillimetre sources are weak sources, some of which are found within extended SIMBA emission, others of which lie close to a bright millimetre SIMBA source. Poor sky subtraction around these strong SIMBA sources, means that often faint millimetre emission which would be coincident with these submillimetre sources is not seen in the SIMBA maps. These faint submillimetre sources are also often at the detection threshold of SIMBA and consequently we do not see millimetre emission associated with them.

  Each of these categories of submillimetre source are depicted in column 10 of Table \ref{chap:scuba:scubafluxes}, with an `R' in this column indicating those submillimetre cores which resolve a known SIMBA core, an `N' indicating no SIMBA emission associated with the submillimetre core, and an `E' indicating those submillimetre cores which fall within the extended emission of a SIMBA core, but have no direct correlation with a SIMBA source themselves.

   Of the 78 sources targeted in this survey, all but three reveal a SCUBA core at both 450 and 850\microm\, corresponding to the millimetre continuum emission detected by SIMBA. The map of G\,29.969-0.003 displays no evidence of a submillimetre core at the targeted position as a result of chopping nearby emission, which creates negative bowls of emission at the centre of the image. The remaining two maps, G\,49.459-0.317 and G\,49.508-0.409, are devoid of emission at the position of the SIMBA core, but display emission at other points in the image. These two images are weak SIMBA sources at 0.3 and 0.4\,Jy respectively, although that should not entirely explain the lack of emission at the targeted position for the following reasons. First, sources of comparable 1.2\,mm flux and distance are detected by SCUBA in this survey (e.g. G\,29.193-0.073), the submillimetre fluxes are typically of order 3 times brighter than SIMBA at 850\microm, and 30 times brighter than SIMBA at 450\microm \footnote{As determined through comparison of the SCUBA fluxes reported by \citet{walsh03} and \citet{thompson06} and the corresponding millimetre flux reported in Paper I.}, and  lastly, emission is detected in these fields, although not at the targeted position. These two non-detections are not a consequence of pointing/positional error.

   Chopping into emission from a nearby submillimetre source results in negative `bowls' of emission in the SCUBA images (e.g. G\,29.969-0.003, G\,49.528-0.348). Also evident in the images (cf. Fig. \ref{chap:scuba:scubasample}) are `holes', which are indicative of bad/noisy bolometers during the jigglemap scans, which were removed during the data reduction procedure. The submillimetre emission of sources found near to these bad bolometers is likely to be undersampled, and consequently the flux values reported for the sources in Table \ref{chap:scuba:scubafluxes} are likely a lower limit. These sources are denoted by a $^\delta$ in columns 6 or 8 of Table \ref{chap:scuba:scubafluxes}. 

   The resultant map of G\,49.528-0.348 has a negative bowl of emission near to the centre of the image, coincident with the source targeted, and as such a flux density is not measured for this source. Unfortunately during the night of September 20, 2003 (fallback observations), no suitable calibrator was observed at 450\microm, leaving these data for this period uncalibrated.

   The SCUBA maps also reveal two types of source located at the edge of the images. Sources denoted by a $^\gamma$ in Table \ref{chap:scuba:scubafluxes} are situated close to the edge of the map, but far enough away, that their morphology suggests that the source has been reasonably sampled. However, due to the small map sizes offered by SCUBA, it is possible that emission extends off the side of the map. Consequently, the size of the source would be underestimated and the flux undersampled, with flux values reported in Table \ref{chap:scuba:scubafluxes} a lower limit to the actual value. Sources situated too close to the edge of the map, including sources which extend off the map are denoted by a $^\zeta$ in Table \ref{chap:scuba:scubafluxes}. Due to uncertain source sizes, the flux value reported here is a lower limit. For those cores for which it appears that the majority of the core extends of the map, a flux is not estimated.

 \subsection{SCUBA Galactic Plane Map}

  \citet{p-p00} undertook a large scale submillimetre continuum map of the Galactic centre region at 450 and 850\microm\, using the SCUBA instrument on the JCMT. The map size is approximately 2.8\degr\,$\times$\,0.5\degr, or 400\,$\times$\,75 pc at an assumed distance of 8.5 kpc. The observation and data reduction procedure are as described in \citet{p-p00}. The map was made in scan-map mode (see \citealp{holland99}) and was calibrated using Uranus and Mars.

  We have examined the calibrated map (Jy/pixel) for SIMBA sources in order to obtain their equivalent submillimetre fluxes. The flux density of each source in this map was determined using the KARMA/KVIS package, by defining a box aperture around the source and at various points in the image considered to be the background. The flux inside each of the apertures was then measured and the latter was subtracted from the former. Due to an artefact in the map, it was not possible to determine the flux for sources spanning the artefact, such as G\,1.13-0.11 and nearby sources. The peak fluxes of sources appearing in this map were determined according to the peak pixel value of the source, with 3 arcsec pixels, assuming a beam size of 8 and 15 arcsec for the 450 and 850\microm\, maps respectively.

   SIMBA sources for which the submillimetre flux was determined using the Galactic Plane map of \citet{p-p00} are denoted by `GPM' in column 5 of Table \ref{chap:scuba:scubafluxes}.

\subsection{SCUBA in the context of the SIMBA survey}\label{chap:scuba:simscuba}

   The impetus of the SIMBA 1.2\,mm continuum survey on the SEST (Paper I) is of particular importance to this SCUBA submillimetre continuum survey.  The SIMBA survey was targeted toward 131 regions exhibiting signs of massive star formation manifest in the form of methanol maser and/or radio continuum emission. The SIMBA source list was predominantly drawn from the existing SCUBA surveys of \citet{walsh03} and \citet{thompson06} (as described in Paper I), which were respectively targeted toward methanol masers and \uchii\, regions. Consequently, a moderate sample of the SIMBA source list reported in Paper I had existing, complementary submillimetre continuum data prior to the undertaking of the SIMBA survey.

   The SIMBA source list is comprised of four distinct millimetre continuum classes. Class M sources have an associated methanol maser source, class R have an associated radio continuum source, whilst class MR have both a methanol maser and radio continuum source. The fourth class (MM) is comprised of sources previously unknown and devoid of star formation tracers such as a methanol maser and/or a radio continuum source, identified solely from their millimetre continuum emission in the SIMBA survey. Analysis, in Paper I, revealed these mm-only cores to have masses that support stars in excess of 10\mstar\, forming, although it is not yet clear which of the mm-only cores will form massive clusters.  The aim of this SCUBA survey was to obtain the complementary submillimetre data for these mm-only cores. 

   Examination of the SIMBA images for correlations with the SCUBA data of this survey, as presented in Table \ref{chap:scuba:scubafluxes}, revealed evidence of two extra SIMBA sources which were not previously noted in Table 5 of Paper I. These two sources (G\,8.111+0.257  and G\,5.962-1.128) are mm-only sources, that is, they are devoid of methanol maser and radio continuum emission. These two sources are denoted by an asterisk in column 3 of Table \ref{chap:scuba:tab_simscuba}. Examination of the images from the two data sets also revealed two SIMBA sources from Paper I, which were observed on two separate occasions, to be the same source (G\,22.35+0.06 and G\,22.36+0.07B). The peak of the SIMBA emission corresponds to that of G\,22.36+0.07B and as such this is assumed to be the dominant source, with a flux of 2.5\,Jy. Coincidentally, the map of G\,22.36+0.07B is also subject to less noise than that of G\,22.35+0.06. These results bring the total number of sources in the SIMBA survey of Paper I, to 405, and the total number of mm-only cores to 255.

   The combination of the four SCUBA surveys -- \citet{walsh03}, \citet{thompson06}, \citet{p-p00} as well as this one, samples a reasonable portion of the SIMBA source list of 403 reported in Paper I, plus the 2 new cores detected here. 113 of the 405 sources (28 per cent) in the SIMBA source list are below the declination limits of the JCMT, and thus it is not currently possible to obtain the complementary submillimetre data for these sources.

   Of the 292 sources in the SIMBA source list obtainable with the JCMT, and hence submillimetre wavelengths, we have complementary submillimetre data at both 450 and 850\microm\, for 154 of these in total (i.e. 53 per cent). For 17 other SIMBA sources, we have 850\microm\, data to complement the 1.2\,mm emission obtained with SIMBA, however the 450\microm\,  flux could not be determined for these sources from the various SCUBA surveys discussed here. The 154 sources from the SIMBA source list that have complementary submillimetre data have been reproduced here in Table \ref{chap:scuba:tab_simscuba}, with their millimetre fluxes (from Paper I) and corresponding fluxes from the combined SCUBA surveys.

   In a few instances the sources reported in Table \ref{chap:scuba:scubafluxes} have also been observed in the surveys of \citet{walsh03} or \citet{thompson06}. For those sources in this survey for which an ambiguity in flux exists, i.e. those sources located near to the edge off the map or near a noisy bolometer, the flux of the \citeauthor{walsh03} or \citeauthor{thompson06} surveys has been adopted for analysis purposes. The survey from which the SCUBA data is drawn is indicated in column 9 of Table \ref{chap:scuba:tab_simscuba}.

\section{Determining the dust emissivity exponent -- $\beta$}\label{chap:scuba:beta_exp}

   Using the millimetre and submillimetre continuum data collated from the SIMBA and SCUBA surveys, and assuming that the 1.2\,mm continuum emission detected toward these regions of massive star formation is from optically thin dust, then the following relation can be used to determine the dust grain emissivity index:

\begin{equation}
F_{\nu} = \Omega_c~B_\nu(T_{dust})~\epsilon_{\nu}
\label{eqbeta1}
\end{equation}
%can quote goldsmith 97 if necessary
\noindent
where $ F_{\nu}$ is the flux density of the source, $\Omega_c$ is the source solid angle, $B_\nu$ is the Planck function for a temperature $T_{dust}$, and $\epsilon_{\nu}$ is the emissivity and is equivalent to (1\,-\,e$^{-\tau_{\nu}}$).

   For optically thin regions, $\epsilon_{\nu}$ approximates to $\tau_{\nu}$ where $\tau_{\nu}$ is given by $\tau_0(\nu/\nu_0)^{\beta}$. Substituting this into equation \ref{eqbeta1} then gives:

\begin{equation}
F_{\nu} =\Omega_c \frac{2h\nu^3}{c^2~(e^{h\nu/kT}-1)} \tau_0(\nu/\nu_0)^{\beta}
\label{eqbeta2}
\end{equation}

\noindent
where $\tau_0$ is the optical depth at which the material becomes optically thin (i.e., $\tau_0 \sim$\,1), $\nu$ is the frequency of observation, $\nu_0$ is the critical frequency at which the material becomes optically thin, and $\beta$ is the dust grain emissivity exponent.

If we collate the constants in equation \ref{eqbeta2} into `A', then Equation \ref{eqbeta2} becomes

\begin{equation}
F_{\nu} = \frac{{\rm A}}{\lambda^{(3+\beta)} (e^{hc/\lambda kT}-1)}
\label{eqbeta_fit}
\end{equation}

\noindent
where A =2hc$\Omega_c \lambda_0^{\beta} \tau_0$.

   Using the relation described in Equation \ref{eqbeta_fit} the value of the dust grain emissivity index ($\beta$) can be determined for specific cores in the SIMBA source list.

   (Sub)millimetre wavelengths provide a useful diagnostic tool for probing the kinematics of these sources, and in particular the dust grain emissivity index ($\beta$). The SIMBA and SCUBA data combined provides a baseline of three data points to facilitate a fit to Equation \ref{eqbeta_fit}.

   With three data points (1200, 850 and 450\microm), it is possible to fit for a maximum of two free parameters, which in this instance, are the dust grain emissivity index $\beta$ and the constant `A', which is comprised of unknown parameters. Therefore, it is not possible to simultaneously fit for both the temperature of the core and $\beta$ of the core. In the Rayleigh-Jeans limit, the temperature and $\beta$ are degenerate in the fit, and thus increasing one will consequently result in a decrease of the other. Accordingly, we should expect the emissivity index to display variation with a change in temperature.

   As a result of the two free parameter restriction, we have fit the data with Equation \ref{eqbeta_fit}, using a Levenberg-Marquardt least squares fit, for a range of temperatures characteristic of cold cores - 10\,-\,50\,K, in order to determine both $\beta$ and the constant `A'. We anticipate that assuming a temperature across the four types of source in our source list  will have small influence on the resultant data for the fit, and investigate this assumption afterwards. In order to disentangle $\beta$ from temperature a further data point, or knowledge of all components of the constant `A', is required. 

   The dust emissivity exponent $\beta$ has been determined for 154 cores for the aforementioned temperature range. The corresponding value of $\beta$ for each temperature (10\,-\,50\,K) is reported in columns 10 through 14 of Table \ref{chap:scuba:tab_simscuba}. Histograms of $\beta$ with respect to temperature for the total sample are presented in Fig \ref{chap:scuba:hist_temperature}, whilst the histograms of $\beta$ per individual class types are presented in Fig \ref{chap:scuba:hist_tracer}. We examine the value of $\beta$ over the entire sample, as well as over the different types of cores in Section \ref{chap:scuba:analysis}.

\section[]{Presentation of the Data}

\subsection{SCUBA Images}\label{chap:scuba:sampleimages}

\begin{figure*}
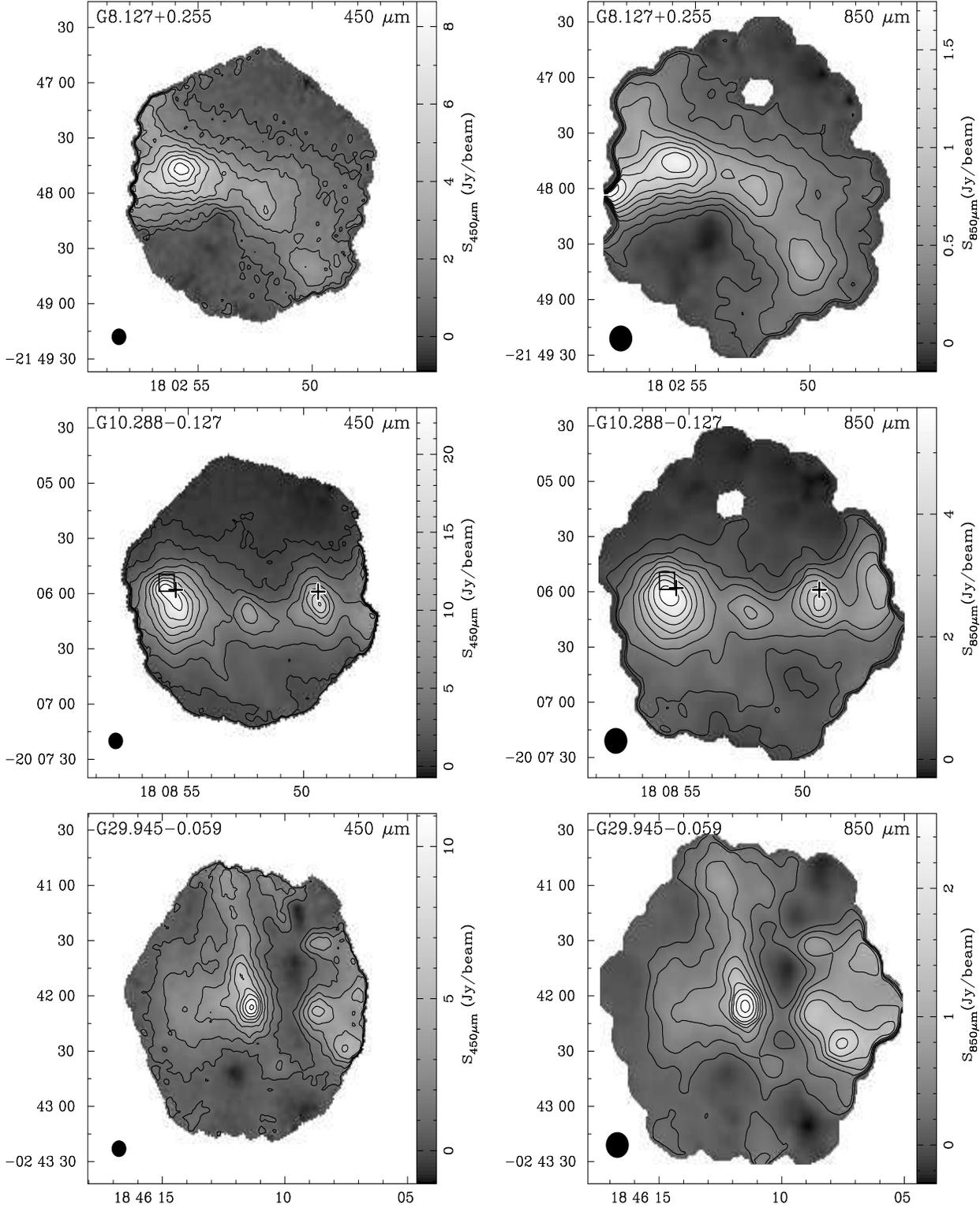

  \begin{tabular}{cc}
         \includegraphics[width=6.6cm, height=8cm, angle=270]{images/G8.127+0.255_sho.ps}&
	 \includegraphics[width=6.6cm, height=8cm, angle=270]{images/G8.127+0.255_lon.ps}\\
         \includegraphics[width=6.6cm, height=8cm, angle=270]{images/G10.288-0.127_sho.ps}&
	 \includegraphics[width=6.6cm, height=8cm, angle=270]{images/G10.288-0.127_lon.ps}\\
         \includegraphics[width=6.6cm, height=8cm, angle=270]{images/G29.945-0.059_sho.ps}&
	 \includegraphics[width=6.6cm, height=8cm, angle=270]{images/G29.945-0.059_lon.ps}\\
  \end{tabular}
\caption{Sample images of the 450 and 850\microm\, emission measured from the SCUBA survey. The `plus' symbol represents the position of methanol maser sources, and the `box' indicates radio continuum sources (\uchii\, region) which fall within the maps observed. The source at the centre of the image, is the mm-only source of the SIMBA survey which was targeted here, with the source name depicted in the top left corner of the image. Contour levels are as described in the text (see Section \ref{chap:scuba:sampleimages}). `Holes' in the images are indicative of bad/noisy bolometers during the jigglemap scans, which have been removed in the data reduction process. The beam size at each wavelength is indicated in the lower left corner of each image. The entire image set from this survey can be found in the Appendix \ref{chap:scuba:scubaimages}.
\label{chap:scuba:scubasample}}
\end{figure*}

   In Figure \ref{chap:scuba:scubasample} we present the 450 and 850\microm\, images for a sample of three sources of the SCUBA survey. Methanol maser and radio continuum sources falling within the maps are indicated by `plus' and `box' symbols respectively. Sources found nearest the centre of the frame are the mm-only cores targeted in this survey. All of the images from this SCUBA survey, including those in this sample set, can be found in the Appendix, Figure \ref{chap:scuba:scubaimages}. Unfortunately, the 850\microm\, map of G\,11.948-0.003 is subject to an artefact running across the image. Consequently, the 850\microm\, flux for this source is not reported, nor is the map shown here. As a result of a negative bowl of emission in the map of G\,49.528-0.308, it was not possible to obtain the flux measurements for this source, and consequently, we do not include the image of this source here. Two sources of the sample, G\,11.903-1.40 and G\,49.456-0.354 have been observed during both the 2003 and 2005 period. The maps of these sources from the May observing period are presented in Figure \ref{chap:scuba:scubaimages}.
   
The flux contour levels on each image were determined by a dynamic range power-law fitting scheme in order to emphasise both the low-level and bright emission. Logarithmic contour levels were fitted to the dynamic range of the image $D$ (defined as the image peak divided by the 1$\sigma$ r.m.s.~noise) following the relation $D = 3 \times N^{i} +3$, where $N$ is the number of contours (in this case 9) and $i$ is the contour power-law index. The minimum power-law index used was one, which results in linear contours starting at a level of 3$\sigma$ and spaced by 3$\sigma$. This dynamic power-law contouring scheme was found to give excellent results in both high and low dynamic range images, concentrating the contours around low surface brightness features in the images to adequately represent low-level structure in the images.

\subsection{Physical Parameters of the Star Forming Regions}

\subsubsection{SCUBA results}

   We present the results of this survey in Table \ref{chap:scuba:scubafluxes}. The table lists all of the sources detected in this survey in right ascension and declination order, using J2000 epoch. Column 3 lists the name of the source in G-name nomenclature, as per the SIMBA survey (Paper I). Galactic names are given to three decimal places in order to distinguish closely associated sources. Column 4 indicates the type of source, consistent with Paper I, where `{\it m}', `{\it r}', `{\it mm}' or `{\it submm}' are used to indicate a maser, radio continuum, mm-only, or submillimetre source, respectively. Column 5 lists the map in which the source was found if not specifically targeted in this survey. A `GPM' in this column indicates sources that were found in the Galactic Plane map of \citet{p-p00}.  Columns 6 and 7 list the integrated (Jy) and peak (Jy/beam) flux of the source at 450\microm, while columns 8 and 9 give the integrated and peak fluxes for the 850\microm\, data. In column 10, the SIMBA associations for each of the submillimetre cores detected in this survey are given. An `R' indicates submillimetre sources which resolve a known SIMBA source, an `N' indicates submillimetre sources devoid of 1.2\,mm emission and hence no SIMBA association, and an `E' indicates those submillimetre cores which fall within the extended emission of a SIMBA source, but have no direct correlation with a SIMBA source themselves.

\begin{table*}
%table=scubafluxes.sxc
  \begin{center}
    \caption{Parameters for the 212 submillimetre continuum sources found in the SCUBA survey. \label{chap:scuba:scubafluxes}}
    \vspace{-0.5cm}

  \end{center}
  \begin{tabular}{@{}lllclccccc@{}}
     \hline

    \multicolumn{2}{c}{Peak Position} & & \multicolumn{2}{c}{Identifier} & \multicolumn{2}{c}{450\microm~ Flux} & \multicolumn{2}{c}{850\microm~ Flux}& \\
    RA & Dec              & Source Name & Tracer & SIMBA target          &  Integ. & Peak &  Integ. & Peak & SIMBA\\
    (J2000)&(J2000)       &   $^a$      &        &                       &  Jy$^b$& Jy/bm &   Jy$^b$& Jy/bm & corr$^c$ \\
(1)&(2)&(3)&(4)&(5)&(6)&(7)&(8)&(9) & (10)\\
\hline
17 45 54.3 & -28 44 08 & G0.204+0.051 & mm & GPM & 33.3	 & 9.5 & 3.7	 & 3.2 &\\
17 46 03.9 & -28 24 58 & G0.49+0.19 & m & G0.497+0.170 & 17.4$^\zeta$ & 12.9 & 2.7$^\delta$ & 1.7 &\\
17 46 07.1 & -28 41 28 & G0.266-0.034 & mm & GPM & 73.6	 & 19.2 & 7.3	 & 5.9 &\\
17 46 07.7 & -28 45 20 & G0.21-0.00 & mr & GPM & 103.6	 & 13.7 & 6.9	 & 4.4 &\\
17 46 07.7 & -28 41 40 & G0.264+0.031 & submm & G0.27-0.03 & 37.2	 & 2.6 & 3.3	 & 5.9 & E\\
17 46 08.2 & -28 25 23 & G0.497+0.170 & mm &   & 15.6	 & 3.9 & 5.8	 & 1.0 &\\
17 46 08.9 & -28 43 44 & G0.237+0.009 & submm & G0.240+0.008 & no cal	 & no cal & 26.9$^\alpha$ & 6.5 & R\\
17 46 09.2 & -28 43 37 & G0.247+0.012 & submm & G0.240+0.008 & no cal	 & no cal & -$^\alpha$ & 6.5 & R\\
17 46 09.4 & -28 24 51 & G0.507+0.171 & submm & G0.497+0.170 & 7.6	 & 3.9 & 2.7$^\delta$	 & 0.8 &E\\
17 46 09.5 & -28 43 36 & G0.240+0.008 & mm & GPM & 216.9	 & 19.3 & 25.5	 & 6.6 &\\
17 46 09.6 & -28 23 46 & G0.523+0.180 & submm & G0.527+0.181 & -$^\alpha$ & 16.9 & -$^\alpha$ & - & R\\
17 46 09.8 & -28 43 10 & G0.247+0.011 & submm & G0.240+0.008 & no cal	 & no cal & -$^\delta$ & 7.0 & E\\
17 46 10.1 & -28 23 31 & G0.527+0.181 & r &   & 45.9$^\alpha$ & 23.2 & 7.6$^\alpha$ & 3.0 &\\
17 46 10.7 & -28 41 36 & G0.271+0.022 & mm & GPM & 17.2	 & 12.5 & 1.9	 & 1.0 &\\
17 46 10.6 & -28 43 29 & G0.244+0.006 & submm & G0.240+0.008 & no cal	 & no cal & 5.1	 & 5.8 & E\\
17 46 11.2 & -28 23 45 & G0.526+0.175 & submm & G0.527+0.181 & 12.3	 & 8.9 & 1.4	 & - & R\\
17 46 11.4 & -28 42 40 & G0.257+0.011 & mm & GPM & 251.1	 & 21.9 & 23.1	 & 7 &\\
17 46 52.8 & -28 07 35 & G0.83+0.18 & m & GPM & ND	 & - & 3.9	 & 2.3 &\\
17 46 58.1 & -28 45 21 & G0.308-0.158 & submm & G0.331-0.164 & no cal	 & no cal & -$^\alpha$ & 0.5 &R\\
17 46 58.9 & -28 45 25 & G0.308-0.161 & submm & G0.331-0.164 & no cal	 & no cal & -$^\alpha$ & 0.6 &R\\
17 46 59.5 & -28 45 34 & G0.307-0.165 & submm & G0.331-0.164 & no cal	 & no cal & 1.3$^\alpha$ & 0.6 &R\\
17 47 01.2 & -28 45 36 & G0.310-0.170 & mm & GPM & 8.8	 & 6 & 0.9	 & 2.3 &\\
17 47 09.1 & -28 46 16 & G0.32-0.20 & mr & GPM & 186.8	 & 15.4 & 18.8	 & 2.2 &\\
17 47 20.1 & -28 47 04 & G0.325-0.242 & mm & GPM & ND	 & - & 1.8	 & 1.8 &\\
17 48 31.6 & -28 00 31 & G1.124-0.065 & mm & GPM & ND	 & - & 2.3	 & 1.5 &\\
17 48 34.7 & -28 00 16 & G1.134-0.073 & mm & GPM & ND	 & - & 0.9	 & 0.8 &\\
17 48 36.4 & -28 02 31 & G1.105-0.098 & mm & GPM & 37.3	 & - & 5.8	 & 2.3 &\\
17 48 41.9 & -28 01 44 & G1.13-0.11 & r & GPM & -$^\epsilon$ & - & -$^\epsilon$ & - &\\
17 48 48.5 & -28 01 13  & G1.14-0.12 & m & GPM & -$^\epsilon$ & - & -$^\epsilon$ & - &\\
17 50 17.6 & -28 53 18 & G0.569-0.851 & submm & G0.549-0.868 & -$^\alpha$ & 6.3 & -$^\alpha$ & - &R\\
17 50 17.6 & -28 53 30 & G0.567-0.853 & submm & G0.549-0.868 & -$^\alpha$ & 6.8 & -$^\alpha$ & - &R\\
17 50 18.3 & -28 53 22 & G0.570-0.853 & submm & G0.549-0.868 & -$^\alpha$ & 6.3 & -$^\alpha$ & - &R\\
17 50 18.8 & -28 53 14 & G0.549-0.868 & mm &   & 15.4$^\alpha$ & 6.4 & 2.3$^\alpha$ & 1.6 &\\
17 50 20.4 & -28 53 47 & G0.568-0.864 & submm & G0.549-0.868 & 8.6	 & 5.5 & 1.3	 & 0.9 &N\\
17 50 23.9 & -28 53 32 & G0.578-0.873 & submm & G0.600-0.871 & ND	 & - & 1.1$^\gamma$ & 0.4 &N\\
17 50 23.9 & -28 50 15 & G0.627-0.848 & mm &   & 12.9$^\alpha$ & 6.6 & 3.3$^\alpha$ & 1.2 &\\
17 50 24.4 & -28 50 40 & G0.620-0.850 & submm & G0.627-0.848 & -$^\alpha$ & 3.9 & -$^\alpha$ & - & R\\
17 50 26.7 & -28 50 56 & G0.620-0.859 & submm & G0.627-0.848 & ND	 & - & 0.9	 & 0.5 &N \\
17 50 26.7 & -28 52 23 & G0.600-0.871 & mm &   & 20.9$^\alpha$ & 3.3 & 3.4$^\alpha$ & 1.2 &\\
17 50 27.8 & -28 52 08 & G0.605-0.873 & submm & G0.600-0.871 & -$^\alpha$ & - & -$^\alpha$ & 0.4 &R\\
17 50 29.7 & -28 51 49 & G0.613-0.876 & submm & G0.600-0.871 & ND	 & - & -$^\alpha$ & 0.4 &E\\
17 50 30.0 & -28 51 58 & G0.612-0.878 & submm & G0.600-0.871 & ND	 & - & 0.4$^\alpha$ & 0.4&E \\
17 59 03.4 & -24 19 24 & G5.948-0.233 & submm & G5.504-0.246 & 0.9	 & 1.4 & 0.2	 & 0.5 &N\\
17 59 04.4 & -24 19 29 & G5.499-0.237 & submm & G5.504-0.246 & 1.1	 & 1.5 & 0.5	 & 0.5 &N\\
17 59 06.0 & -24 19 28 & G5.502-0.242 & submm & G5.504-0.246 & -$^\alpha$ & 1.9 & -$^\alpha$ & 0.7 &R\\
17 59 07.5 & -24 19 19  & G5.504-0.246 & mm &   & 17.2$^\alpha$ & 2.3 & 2.8$^\alpha$ & 0.9 &\\
17 59 07.6 & -24 19 14 & G5.509-0.246 & submm & G5.504-0.246 & -$^\alpha$ & 2.3 & -$^\alpha$ & - &R\\
17 59 09.4 & -24 19 05 & G5.514-0.250 & submm & G5.504-0.246 & 0.7	 & 1.2 & 0.4	 & 0.4 &N\\
% &  &  &  &  & 	 &  & 	 &  &\\
18 02 50.2 & -21 48 39 & G8.111+0.257 & mm & G8.127+0.255 & 6.3$^\alpha$ & 2.7 & 0.6$^\alpha$ & 0.7 &\\
18 02 51.9 & -21 48 08 & G8.122+0.256 & submm & G8.127+0.255 & -$^\alpha$ & 3.7 & -$^\alpha$ & - &R\\
18 02 52.8 & -21 47 54 & G8.127+0.255 & mm &   & 15.7$^\alpha$ & 3.5 & 1.9$^\alpha$ & 0.8 &\\
18 02 56.2 & -21 47 38 & G8.138+0.246 & mm &   & 38.7	 & 8.6 & 4.0	 & 1.6 &\\
18 02 58.4 & -21 48 01 & G8.136+0.235 & submm & G8.138+0.246 & 6.0	 & 6.1 & 2.4$^\alpha$ & 1.7 &E\\
18 02 59.2 & -21 48 06 & G8.137+0.232 & submm & G8.138+0.246 & 5.9	 & 6.1 & -$^\alpha$ & 1.7 &E\\
18 02 59.8 & -21 47 48 & G8.142+0.232 & submm & G8.138+0.246 & 4.0	 & 5.3 & -$^\phi$ & - &E\\
18 03 26.3 & -24 22 29 & G5.948-1.125 & mm &   & 7.2	 & 2.3 & 1.0	 & 0.7 &\\
18 03 28.7 & -24 21 50 & G5.962-1.128 & mm & G5.948-1.125 & 5.5$^\gamma$ & 2.2 & 1.7$^\gamma$ & 0.6 &\\
18 03 33.3 & -24 21 46 & G5.971-1.143 & submm & G5.975-1.146 & 7.1	 & 3.8 & -$^\alpha$ & - &R\\
18 03 34.5 & -24 21 41 & G5.975-1.146 & mm &   & 4.3	 & 3.9 & 1.3$^\alpha$ & 1.0 &\\
18 03 35.7 & -24 22 05 & G5.971-1.153 & submm & G5.975-1.146 & -$^\alpha$ & 5.7 & -$^\alpha$ & - & R\\
18 03 36.8 & -24 22 13 & G5.971-1.158 & mm & G5.975-1.146 & 27.4$^\alpha$ & 7.5 & 4.3$^\alpha$ & 1.8& \\
\end{tabular}
\end{table*}

\begin{table*}
\contcaption{}
\vspace{-0.3cm}
    \begin{tabular}{@{}lllclccccc@{}}
    \hline
    \multicolumn{2}{c}{Peak Position} & & \multicolumn{2}{c}{Identifier} & \multicolumn{2}{c}{450\microm~ Flux} & \multicolumn{2}{c}{850\microm~ Flux}&\\
    RA & Dec              & Source Name & Tracer & SIMBA target          &  Integ. & Peak &  Integ. & Peak & SIMBA\\
    (J2000)&(J2000)       &   $^a$      &        &                       &  Jy$^b$& Jy/bm &   Jy$^b$& Jy/bm & corr$^c$ \\
(1)&(2)&(3)&(4)&(5)&(6)&(7)&(8)&(9)&(10) \\
\hline
18 03 37.4 & -24 22 25 & G5.970-1.161 & submm & G5.975-1.146 & -$^\alpha$ & 5.1 & -$^\alpha$ & - &R\\
18 08 45.9 & -20 05 34 & G10.287-0.110 & mm &   & 51.0$^\alpha$ & 7.1 & 14.4$^\alpha$ & 3.2 &\\
18 08 45.8 & -20 05 51 & G10.282-0.112 & submm & G10.287-0.110 & -$^\alpha$ & 5.6 & -$^\alpha$ & - &R\\
18 08 47.3 & -20 06 10 & G10.281-0.120 & submm & G10.287-0.110 & -$^\alpha$ & 5.4 & -$^\alpha$ & - &R\\
18 08 49.4 & -20 05 58 & G10.284-0.126 & m & G10.288-0.127 & 53.9	 & 16.7 & 9.8	 & 4.0 &\\
18 08 52.4 & -20 05 58 & G10.288-0.127 & mm &   & 30.6	 & 9.1 & 5.0	 & 2.3 &\\
18 08 55.5 & -20 05 58 & G10.29-0.14 & mr & G10.288-0.127 & 148.2$^\gamma$ & 23.0 & 29.0	 & 5.7 &\\
18 09 00.0 & -20 03 34 & G10.343-0.142 & m & G10.359-0.149 & -$^\zeta$ & - & 11.7$^\gamma$ & 4.5 &\\
18 09 03.5 & -20 02 54 & G10.359-0.149 & mm &   & 52.2	 & 8.2 & 4.6	 & 1.7 &\\
18 09 06.0 & -20 03 56 & G10.349-0.166 & submm & G10.359-0.149 & ND	 & - & 0.4	 & 0.8 &N\\
18 10 15.7 & -19 54 45 & G10.63-0.33B & mm &   & 18.3	 & 7.2 & 4.0	 & 1.5 &\\
18 10 17.0 & -19 54 54 & G10.615-0.336 & submm & G10.63-0.33B & 8.9	 & 6.0 & 1.0	 & 1.0 &R\\
18 10 18.0 & -19 54 05 & G10.62-0.33 & m & G10.63-0.33B & 67.0$^\zeta $ & 11.3 & 12.7$^\delta$ & 2.2 &\\
18 10 24.1 & -20 43 09 & G9.924-0.749 & mm &   & no cal	 & no cal & 1.7	 & 1.3 &\\
18 11 24.4 & -19 32 04 & G11.075-0.384 & mm &   & no cal	 & no cal & 3.0	 & 1.4 &\\
18 11 52.9 & -18 36 03 & G11.948-0.003 & mm &   & no cal	 & no cal & -$^\epsilon$ & - &\\
18 11 53.6 & -17 30 02 & G12.914+0.493 & mm &   & 13.7	 & 6.2 & 1.8	 & 1.6 &\\
18 11 56.9 & -17 29 35 & G12.927+0.485 & submm & G12.914+0.493 & 0.8	 & 1.1 & 0.2	 & 0.3 &N\\
18 12 11.1 & -18 41 27 & G11.903-0.140 & mr &   & 35.0	 & 13.6 & 3.7	 & 3.3 &\\
18 12 14.7 & -18 45 06 & G11.857-0.181 & submm & G11.861-0.183 & -$^\alpha$ & - & -$^\alpha$ & 0.3 &R\\
18 12 15.6 & -18 44 58 & G11.861-0.183 & mm &   & 6.5$^\alpha$ & 1.9 & 0.6$^\alpha$ & 0.3 &\\
18 12 16.3 & -18 44 58 & G11.862-0.186 & submm & G11.861-0.183 & -$^\alpha$ & - & -$^\alpha$ & 0.3 &R\\
18 12 17.3 & -18 40 03 & G11.93-0.14 & m & G11.942-0.157 & 17.6	 & 8.0 & 2.1	 & 1.6 &\\
18 12 19.2 & -18 40 20 & G11.935-0.159 & submm & G11.942-0.157 & 1.5	 & 2.4 & 0.3	 & 0.7 &E\\
18 12 19.6 & -18 39 54 & G11.942-0.157 & mm &   & 16.2	 & 10.6 & 1.7	 & 2.1 &\\
18 12 39.2 & -18 24 17 & G12.20-0.09 & mr & G12.216-0.119 & -$^\zeta$ & 53.2 & 14.8$^\zeta$ & 10.6 &\\
18 12 42.7 & -18 25 08 & G12.18-0.12 & m & G12.216-0.119 & 11.5	 & 6.6 & 4.1	 & 2.3 &\\
18 12 44.5 & -18 25 09 & G12.205-0.125 & submm & G12.216-0.119 & 11.0	 & 5.5 & 3.1	 & 1.7 & R\\
18 12 44.4 & -18 24 25 & G12.216-0.119 & mm &   & 37.9	 & 15.1 & 5.9	 & 1.3 &\\
18 13 58.5 & -18 54 21 & G11.94-0.62B & mm &   & 66.1	 & 15.5 & 14.0	 & 4.3 &\\
18 14 00.9 & -18 53 27 & G11.93-0.61 & mr & G11.94-0.62B & ND	 & - & -$^{\zeta \delta}$ & 5.1 &\\
18 14 07.6 & -18 00 37 & G12.722-0.218 & mm &   & 25.7	 & 13.1 & 5.9	 & 2.9 &\\
18 14 7.18 & -18 01 28 & G12.709-0.223 & submm & G12.722-0.218 & 3.3	 & 2.6 & 0.6	 & 0.8 &N\\
18 14 34.3 & -17 51 56 & G12.90-0.25B & mm &   & 34.9	 & 17.1 & 4.2	 & 2.9 &\\
18 14 36.1 & -17 54 56 & G12.859-0.272 & mm &   & 47.7	 & 24.3 & 5.4	 & 3.3 &\\
18 21 14.7 & -14 33 38 & G16.569-0.085 & submm & G16.580-0.079 & -$^\phi$ & 0.3 & 0.4	 & 0.5 &E\\
18 21 14.9 & -14 33 04 & G16.578-0.081 & submm &   & 2.4	 & 0.9 & 1.8$^\alpha$ & 0.9 &R\\
18 21 15.8 & -14 32 51 & G16.583-0.083 & submm & G16.580-0.079 & 0.9	 & 0.4 & -$^\alpha$ & - &R\\
18 25 05.1 & -13 14 55 & G18.165-0.293 & mm & G18.177-0.296 & -$^\phi$ & 4.6 & -$^\delta$ & - &\\
18 25 07.3 & -13 14 23 & G18.177-0.296 & mm &   & 29.7	 & 10.8 & 2.8	 & 2.2 &\\
18 27 38.2 & -11 56 38 & G19.607-0.234 & mr &   & 190.1	 & 99.4 & 26.9	 & 20.1& \\
18 29 24.2 & -15 15 34 & G16.871-2.154 & mm &   & 164.7	 & 20.0 & 30.3	 & 6.3 &\\
18 29 24.4 & -15 16 04 & G16.86-2.15 & m & G16.871-2.154 & 101.7	 & 29.2 & 27.9	 & 8.4 &\\
18 29 30.9 & -15 15 52 & G16.879-2.180 & submm & G16.883-2.188 & 0.9	 & 2.1 & 0.1	 & 0.4 &N\\
18 29 32.7 & -15 15 49 & G16.883-2.188 & submm &   & 6.0	 & 3.3 & 0.5	 & 0.8 &R\\
18 29 34.2 & -15 15 44 & G16.887-2.191 & submm & G16.883-2.188 & 6.4	 & 3.0 & 1.2	 & 0.8 &R\\
18 33 52.3 & -08 08 20 & G23.692+0.167 & submm & G23.689+0.159 & -$^\alpha$ & 2.8 & -$^\alpha$ & - &E\\
18 33 52.9 & -08 08 15 & G23.695+0.165 & submm & G23.689+0.159 & -$^\alpha$ & 2.6 & -$^\alpha$ & - &E\\
18 33 53.5 & -08 08 48 & G23.688+0.158 & submm &   & 6.3	 & 3.6 & 0.8	 & 0.6 &R\\
18 33 53.9 & -08 08 21 & G23.695+0.161 & submm & G23.689+0.159 & 7.4$^\alpha$ & 3.0 & 0.6$^\alpha$ & 0.6 &E\\
18 33 54.7 & -08 08 28 & G23.695+0.157 & submm & G23.689+0.159 & 6.7	 & 3.5 & 0.6	 & 0.7 &R\\
18 33 56.4 & -08 08 49 & G23.693+0.148 & submm & G23.689+0.159 & 0.4	 & 2.0 & -$^\alpha$ & - &N\\
18 33 57.2 & -08 08 50 & G23.694+0.145 & submm & G23.689+0.159 & 0.6	 & 2.2 & 0.1$^\alpha$ & 0.4 &N\\
18 34 31.3 & -08 42 47 & G23.25-0.24 & m & G23.268-0.257 & -$^\zeta$ & - & 1.9	 & 1.6 &\\
18 34 36.2 & -08 42 39 & G23.268-0.257 & mm &   & 85.3	 & 25.6 & 7.1	 & 2.5 &\\
18 34 50.3 & -08 40 53 & G23.321-0.295 & submm & G23.319-0.298 & ND & - & -$^\alpha$ & 1.3 &R\\
18 24 50.5 & -08 41 09 & G22.172+1.892 & submm & G23.319-0.298 & ND & - & 5.6$^\alpha$ & 1.1& R\\
18 34 51.9 & -08 40 56 & G23.323-0.302 & submm & G23.319-0.298 & ND & - & -$^\alpha$ & 1.2 &R\\
18 34 46.3 & -08 41 30 & G23.304-0.286 & submm & G23.319-0.298 & ND & - & 0.5	 & 0.8 &N\\
18 34 47.3 & -08 41 43 & G23.303-0.291 & submm & G23.319-0.298 & ND & - & 0.2	 & 0.8 &N\\
18 36 06.1 & -07 13 47 & G23.754+0.095 & mm &   & 38.5	 & 22.7 & 3.0	 & 2.3 &\\
18 36 25.9 & -07 05 08 & G24.919+0.088 & mm &   & 72.1	 & 51.1 & 7.0	 & 3.1 &\\
18 42 54.9 & -04 07 40 & G28.287+0.010 & mm &   & 13.3	 & 3.4 & 1.7	 & 0.8 &\\
\end{tabular}
\end{table*}

\begin{table*}
\contcaption{}
\vspace{-0.3cm}
    \begin{tabular}{@{}lllclccccc@{}}
    \hline
    \multicolumn{2}{c}{Peak Position} & & \multicolumn{2}{c}{Identifier} & \multicolumn{2}{c}{450\microm~ Flux} & \multicolumn{2}{c}{850\microm~ Flux} & \\
    RA & Dec              & Source Name & Tracer & SIMBA target          &  Integ. & Peak &  Integ. & Peak & SIMBA\\
    (J2000)&(J2000)       &   $^a$      &        &                     &  Jy$^b$& Jy/bm &   Jy$^b$& Jy/bm  & corr$^c$\\
(1)&(2)&(3)&(4)&(5)&(6)&(7)&(8)&(9)&(10) \\
\hline
18 42 57.7 & -04 07 12 & G28.300+0.003 & submm & G28.287+0.010 & 2.0	 & 1.1 & 0.3	 & 0.3 &N\\
18 42 59.1 & -04 08 05 & G28.289-0.008 & submm & G28.287+0.010 & ND	 & - & 0.1	 & 0.3 &N\\
18 43 02.4 & -04 14 59 & G29.193-0.073 & mm &   & 8.9	 & 3.6 & 1.3	 & 0.9 &\\
18 45 52.8 & -02 42 29 & G29.888+0.001 & mm &   & 14.2$^\alpha$ & 2.9 & 2.8$^\alpha$ & 0.8 &\\
18 45 53.0  & -02 42 29 & G29.889+0.000 & submm & G29.888+0.001 & -$^\alpha$ & 2.0 & -$^\alpha$ & 0.6 &R\\
18 45 54.4 & -02 42 37 & G29.889-0.006 & m  & G29.888+0.001 & 4.7	 & 2.6 & 1.2	 & 0.6 &\\
18 45 58.9 & -02 40 38 & G29.927-0.008 & submm & G29.918-0.014 & 6.2	 & 2.6 & 0.3	 & 0.6 &E\\
18 45 59.0 & -02 41 08 & G29.920-0.012 & submm & G29.918-0.014 & 5.0	 & 3.7 & -$^\alpha$ & - &R\\
18 45 59.9 & -02 41 16 & G29.920-0.016 & submm &   & 5.1	 & 3.6 & 1.2$^\alpha$ & 1.0 &R\\
18 46 00.2 & -02 45 09 & G29.86-0.04 & m & G29.861-0.053 & 30.9	 & 10.7 & 5.0	 & 1.9 &\\
18 46 00.7 & -02 41 44 & G29.915-0.023 & submm & G29.918-0.014 & 1.3	 & 1.9 & 0.1	 & 0.3& N\\
18 46 01.3 & -02 45 25 & G29.861-0.053 & mm &   & 19.7	 & 4.6 & 2.2	 & 1.0 &\\
18 46 02.4 & -02 45 57 & G29.853-0.062 & mm & G29.861-0.053 & 25.6	 & 6.6 & 3.2	 & 1.5 &\\
18 46 05.0 & -02 42 29 & G29.912-0.045 & m &   & 65.4	 & 14.7 & 8.8	 & 3.0 &\\
18 46 06.1 & -02 41 25 & G29.930-0.040 & mm &   & 12.3	 & 7.1 & 3.6	 & 1.3 &\\
18 46 07.6 & -02 42 25 & G29.918-0.054 & submm & G29.912-0.045 & 14.8	 & 9.6 & 3.6	 & 1.5 &R\\
18 46 08.4 & -02 41 31 & G29.932-0.050 & submm & G29.945-0.059 & 1.0	 & 3.6 & 0.2	 & 0.8 &N\\
18 46 08.5 & -02 42 04 & G29.924-0.054 & submm & G29.912-0.045 & 14.6	 & 7.0 & 2.7	 & 1.3 &R\\
18 46 08.8 & -02 39 09 & G29.969-0.033 & mm &   & ND	 & - & ND	 & - &\\
18 46 09.5 & -02 40 56 & G29.943-0.049 & submm & G29.937-0.054 & 1.9	 & 4.8 & 0.3	 & 1.1 &N\\
18 46 09.8 & -02 41 25 & G29.937-0.054 & mm &   & 11.8	 & 8.3 & 2.1	 & 1.3 &\\
18 46 10.2 & -02 42 37 & G29.919-0.065 & submm & G29.945-0.059 & 0.6	 & 1.9 & 0.1	 & 0.5 &N\\
18 46 10.7 & -02 41 02 & G29.944-0.055 & submm & G29.945-0.059 & ND	 & - & 0.2	 & 0.5 &N\\
18 46 11.5 & -02 42 05 & G29.945-0.059 & mm &   & 30.0$^\alpha$ & 11.1 & 3.4$^\alpha$ & 2.6 &\\
18 46 11.6 & -02 41 47 & G29.935-0.064 & submm & G29.945-0.059 & -$^\alpha$ & 6.5 & -$^\alpha$ & 0.8 &R\\
18 46 12.4 & -02 40 59 & G29.948-0.060 & submm & G29.945-0.059 & -$^\phi$ & - & 1.7	 & 0.8 &E\\
18 46 12.5 & -02 39 09 & G29.978-0.050 & m & G29.969-0.033 & -	 & 12.4 & 7.1$^\zeta$ & 2.0 &\\
18 47 09.5 & -01 45 11 & G30.884+0.153 & submm & G30.894+0.140 & 2.0	 & - & 0.7	 & 0.8 &N\\
18 47 11.7 & -01 45 41 & G30.881+0.141 & submm & G30.894+0.140 & 2.5	 & 2.0 & 0.6	 & 0.7 &N\\
18 47 12.5 & -01 44 17 & G30.903+0.148 & submm & G30.894+0.140 & 1.6	 & 2.4 & 0.4	 & 0.6 &N\\
18 47 13.4 & -01 44 58 & G30.894+0.140 & mm &   & 34.6	 & 5.1 & 2.9	 & 1.5 &\\
18 47 26.7 & -01 44 42 & G30.924+0.092 & mm &   & 12.0	 & 3.5 & 1.1	 & 1.1 &\\
18 47 33.4 & -01 57 17 & G30.750-0.028 & submm & G30.760-0.027 & -$^\eta$ & - & 0.5	 & 1.0 &N\\
18 47 34.2 & -01 56 41 & G30.760-0.027 & mm &   & -$^\eta$ & 26.0 & 3.7	 & 1.8 &\\
18 47 34.5 & -01 57 01 & G30.756-0.030 & submm & G30.760-0.027 & -$^\eta$ & - & 0.6	 & 0.9 &E\\
18 47 35.4 & -02 02 07 & G30.682-0.072 & mr &   & 76.6	 & 29.7 & 5.5	 & 2.8 &\\
18 47 36.8 & -01 58 18 & G30.741-0.049 & submm & G30.740-0.060 & -$^\eta$ & 21.1 & 2.7	 & 2.0 &E\\
18 47 37.9 & -01 57 45 & G30.76-0.05 & mm & G30.740-0.060 & 72.1$^\gamma$ & 46.3 & 6.4	 & 4.5 &\\
18 47 38.7 & -01 59 39 & G30.725-0.066 & submm & G30.729-0.078 & -$^\eta$ & - & 0.5	 & 0.7 &E\\
18 47 39.2 & -01 58 41 & G30.740-0.060 & mm &   & 110.5	 & 32.9 & 8.9	 & 4.2 &\\
18 47 39.2 & -02 00 33 & G30.712-0.074 & submm & G30.729-0.078 & 19.1	 & 28.9 & 1.8	 & 2.1 &E\\
18 47 41.3 & -01 59 45 & G30.729-0.078 & mm &   & 42.5	 & 17.2 & 4.4	 & 1.4 &\\
18 47 41.3  & -02 00 33 & G30.716-0.082 & mm & G30.729-0.078 & 137.6	 & 60.3 & 8.1	 & 4.9 &\\
18 48 01.6 & -01 36 01 & G31.119+0.029 & mm &   & 5.4	 & 5.7 & 0.9	 & 0.6 &\\
18 48 07.2 & -01 28 42 & G31.238+0.640 & submm & G31.256+0.061 & 4.1$^\alpha$ & - & 1.4$^\alpha$ & 0.7 &E\\
18 48 08.4 & -01 28 50 & G31.238+0.059 & submm & G31.256+0.061 & -$^\alpha$ & - & -$^\alpha$ & 0.7 &N\\
18 48 08.2 & -01 28 33 & G31.242+0.062 & submm & G31.256+0.061 & -$^\alpha$ & - & -$^\alpha$ & 0.7 &N\\
18 48 09.7 & -01 27 50 & G31.256+0.061 & m &   & 19.9$^\alpha$ & 5.5 & 4.0$^\alpha$ & 1.0 &\\
18 48 10.2 & -01 27 53 & G31.256+0.059 & submm & G31.256+0.061 & -$^\alpha$ & 5.2 & -$^\alpha$ & - &R\\
18 48 10.5 & -01 28 11 & G31.252+0.056 & submm & G31.256+0.061 & -$^\alpha$ & 5.0 & -$^\alpha$ & - &R\\
18 49 33.1 & -01 29 04  & G31.40-0.26 & r & G31.388-0.266 & 100.3	 & 59.4 & 12.0	 & 6.4 &\\
18 49 34.2 & -01 29 44 & G31.388-0.266 & mm &   & 4.9	 & 5.6 & 0.7	 & 0.7 &\\
18 53 16.0 & +01 15 17 & G35.257+0.165 & submm & G34.256+0.155 & 83.3	 & 60.3 & 13.3	 & 8.2 &R\\
18 53 18.2 & +01 14 57 & G34.256+0.155 & mm &   & 1572.3	 & 359.7 & 165.8	 & 56.1& \\
18 56 01.2 & +02 22 59 & G35.57+0.07 & r & G35.586+0.061 & 43.8$^\gamma$ & 9.8 & 4.3	 & 2.2 &\\
18 56 03.8 & +02 22 30 & G35.573+0.054 & submm & G35.586+0.061 & 3.4	 & 1.7 & 0.5	 & 0.6 &R\\
18 56 03.9 & +02 23 23 & G35.586+0.061 & mm &   & 13.4	 & 4.9 & 1.3	 & 1.1 &\\
18 56 05.5 & +02 22 27 & G35.575+0.048 & mm &   & 7.8	 & 3.4 & 1.2	 & 1.0 &\\
18 56 11.5 & +02 21 51 & G35.578+0.021 & submm & G35.574+0.007 & 1.7	 & 1.9 & 0.2	 & 0.5 &E\\
18 56 13.5 & +02 21 39 & G35.575+0.010 & mm & G35.574+0.007 & -$^\alpha$ & - & -$^\alpha$ & - &\\
18 56 14.0 & +02 21 15 & G35.574+0.007 & mm &   & 28.9$^\alpha$ & 3.3 & 3.3$^\alpha$ & 1.2 &\\
% &  &  &  &  & 	 &  & 	 &  \\
19 00 05.5 & +03 59 21 & G37.468-0.103 & submm & G37.475-0.106 & 5.1	 & 3.9 & 0.6	 & 1.6& E\\
19 00 06.9 & +03 59 39 & G37.475-0.106 & mm &   & 6.1	 & 5.5 & 0.6	 & 1.2 &\\
\end{tabular}
\end{table*}

\begin{table*}
\contcaption{}
\vspace{-0.3cm}
    \begin{tabular}{@{}lllclccccl@{}}
    \hline
    \multicolumn{2}{c}{Peak Position} & & \multicolumn{2}{c}{Identifier} & \multicolumn{2}{c}{450\microm~ Flux} & \multicolumn{2}{c}{850\microm~ Flux}&\\
    RA & Dec              & Source Name & Tracer & SIMBA target          &  Integ. & Peak &  Integ. & Peak & SIMBA\\
    (J2000)&(J2000)       &   $^a$      &        &                      &  Jy$^b$& Jy/bm &   Jy$^b$& Jy/bm & corr$^c$ \\
(1)&(2)&(3)&(4)&(5)&(6)&(7)&(8)&(9)& (10) \\
\hline
19 23 24.9 & +14 30 56 & G49.459-0.317 & mm &   & ND	 & - & ND	 & - &\\
19 23 32.6 & +14 29 52 & G49.456-0.354 & mm &   & 83.8	 & 14.2 & 11.7	 & 3.5 &\\
19 24 34.9 & +14 30 23 & G49.584-0.570 & submm & G49.482-0.355 & 31.9	 & 44.4 & 4.2	 & 6.3 &E\\
19 23 35.9 & +14 31 04 & G49.482-0.355 & mm &   & 40.3	 & 53.4 & 5.1	 & 7.5 &\\
19 23 37.3 & +14 30 11 & G49.471-0.367 & submm & G49.482-0.355 & 139.3	 & 69.6 & 19.7	 & 9.5& R\\
19 23 37.7 & +14 30 26 & G49.476-0.367 & submm & G49.482-0.355 & 88.6$^\alpha$ & 72.1 & -$^\alpha$ & - &R\\
19 23 38.2 & +14 34 16 & G49.533-0.338 & mm &   & 5.6	 & 7.1 & 0.8	 & 0.9 &\\
19 23 38.4 & +14 30 37 & G49.481-0.367 & submm & G49.482-0.355 & 164.9$^\alpha$ & 82.0 & 19.6$^\alpha$ & 10.0 &R\\
19 23 39.6 & +14 33 48  & G49.529-0.347 & submm & G49.533-0.338 & -$^{\alpha \xi}$ & - & -$^{\alpha \xi}$ & 15.5 &R\\
19 23 40.3 & +14 33 49 & G49.528-0.348 & mm & G49.533-0.338 & -$^{\alpha \xi}$ & - & -$^{\alpha \xi}$ & - &\\
19 23 39.8 & +14 31 05 & G49.49-0.37 & m & G49.488-0.385 & 244.9	 & 164.4 & 37.0	 & 41.4 &\\
19 23 42.0 & +14 30 55 & G49.491-0.378 & submm & G49.488-0.385 & 175.7	 & 86.7 & 133.6	 & 13.8 &R\\
19 23 43.1 & +14 30 32 & G49.488-0.385 & m &   & 1154.0	 & 489.9 & 132.8	 & 74.6 &\\
19 23 45.9 & +14 29 44 & G49.481-0.401 & submm & G49.488-0.385 & ND	 & - & 13.9$^\gamma$ & 12.2 &R\\
19 23 47.6  & +14 29 29 & G49.481-0.409 & submm & G49.474-0.420 & 20.4$^\alpha$ & 16.5 & -	 & - &R\\
19 23 47.7 & +14 29 23 & G49.479-0.410 & submm & G49.474-0.420 & -$^\alpha$ & 16.7 & -	 & - &R\\
19 23 48.1 & +14 31 28  & G49.511-0.395 & submm & G49.508-0.409 & 4.0	 & 3.6 & 0.4	 & 0.6 &N\\
19 23 49.2  & +14 28 48 & G49.474-0.420 & mm &   & 50.3	 & 12.3 & 4.9	 & 1.7 &\\
19 23 49.3 & +14 31 29 & G49.513-0.399 & submm & G49.508-0.409 & 2.9	 & 4.2 & 0.4	 & 0.6 &N\\
19 23 49.8 & +14 27 49 & G49.461-0.430 & submm & G49.474-0.420 & ND	 & - & 0.2	 & 0.8 &R\\
19 23 49.9 & +14 28 22 & G49.469-0.426 & submm & G49.474-0.420 & 4.2	 & 9.3 & 0.4	 & 1.1 &R\\
19 23 50.8 & +14 29 52 & G49.494-0.420 & mm & G49.508-0.409 & 17.7	 & 8.0 & 1.5$^\gamma$ & 1.6& \\
19 23 50.8 & +14 30 56 & G49.508-0.409 & mm &   & ND	 & - & ND	 & - &\\
19 43 10.0 & +23 44 59 & G59.794+0.076 & mm &   & 21.1	 & 8.7 & 2.4	 & 1.5 &\\
19 43 11.2 & +23 44 03 & G59.78+0.06 & r & G59.794+0.076 & 132.5$^\zeta$ & 29.5 & 21.0$^\gamma$ & 4.1 &\\

\hline
\end{tabular}
\begin{flushleft}
$^a$ Source names are given to three decimal places in order to distinguish closely associated sources.

$^b$ 450 and 850\microm\, fluxes in Jy. A `no cal' in this column indicates those sources that can not be calibrated at 450\microm. `ND' indicates a non-detection.  $^\alpha$ denotes those sources that have more than one submillimetre flux encompassed within, for which it is not possible to clearly distinguish the individual cores. The flux quoted here is for all components of the core. $^\zeta$denotes sources that are situated quite close to the edge of the map such that their flux is not fully sampled, and thus the flux quoted here is a lower limit value. Those sources which do not have a flux reported are situated too close to the edge of the map, that it is not possible to estimate the size and flux of the core to any level of confidence. $^\gamma$indicates sources that are situated near to the edge of the map, but far enough away that we can be reasonably confident in estimating the source size and flux. $^\delta$denotes sources which are affected by a noisy bolometer in the map. Flux values quoted here are a lower limit. $^\epsilon$indicates those sources for which it is not possible to obtain a flux estimate due to the presence of an artefact in the map. $^\phi$is indicative of maps which are too confused or the emission is too diffuse for a flux value to be estimated. $^\eta$denotes cores which were observed in noisy maps, and hence it is not possibly to obtain a flux estimate (rms noise is $\sim$\,5.0\,Jy at 450\microm). $^\xi$ indicates those sources coincident with a negative bowl of emission and as such a flux measurement can not be obtained.

$^c$ Indicates the SIMBA association for the submillimetre cores detected in this survey. An `R' indicates submillimetre sources which resolve a SIMBA millimetre core into multiple components. An `N' indicates those submillimetre cores without any corresponding millimetre continuum emission associated with the SCUBA emission detected here. An `E' denotes those sources which are located within, generally at the edge of, the extended emission of a nearby SIMBA source, for which there is no corresponding SIMBA core.

\end{flushleft}
\end{table*}

\subsubsection{Combining the SIMBA and SCUBA data}\label{chap:scuba:combining}

   In this section (Table \ref{chap:scuba:tab_simscuba}) we concatenate the results from this SCUBA survey with the SCUBA surveys of \citet{walsh03}, and \citet{thompson06} as well as with fluxes obtained from the SCUBA Galactic Plane map of \citet{p-p00} relative to the SIMBA source list (Table 5, Paper I). The Table includes the millimetre fluxes from Paper I for the sources which have complementary SCUBA data from the combined studies. SIMBA sources which are resolved by SCUBA into multiple components, have the equivalent submillimetre fluxes at 450 and 850\microm\, recorded. In order to obtain the submillimetre equivalent for these sources, the individual core components have been summed together. These sources are denoted as indicated below.

   Table \ref{chap:scuba:tab_simscuba} lists the sources in right ascension (RA) order. Columns 1, 2, 3, 4, and 5 have been reproduced directly from Table 5 of Paper I, with the addition of the two extra SIMBA sources included here. Columns 1 and 2 give the coordinates of the source in right ascension and declination, using J2000 epoch. Column 3 lists the source name, with the extended Galactic names intended to distinguish closely associated sources. The two new mm-only sources identified here, are preceded with an asterisk. Column 4 lists the source identifier, with `{\it m}', `{\it r}', or `{\it mm}' used to indicate a maser, radio continuum, or mm-only source respectively. Columns 5, 6 and 7 list the 1.2\,mm, 850\microm, and 450\microm\, integrated flux for each of the SIMBA cores. Column 8 indicates that the SIMBA source is comprised of multiple SCUBA cores, with $\star$, $\diamond$ and $\dagger$ used to denote 2, 3 and 4 SCUBA cores respectively. The survey from which the submillimetre fluxes were obtained is indicated in column 9, where `T' denotes the survey of \citet{thompson06}, `PP' indicates the Galactic Plane map of \citet{p-p00}, `W' indicates the SCUBA data of \citet{walsh03} and `H' indicates the SCUBA fluxes reported in Table \ref{chap:scuba:scubafluxes}. Columns 10 through 14 report the corresponding dust grain emissivity exponent ($\beta$) for temperatures of 10, 20, 30, 40 and 50\,K.

\begin{table*}
  \begin{center}
%table= all_source_data.sxc
    \caption{171 SIMBA sources with corresponding SCUBA data and the dust grain emissivity index $\beta$. Errors in $\beta$ are typically 20 per cent at 10\,K and 30 per cent for the 20\,-\,50\,K range. \label{chap:scuba:tab_simscuba}}
    \vspace{-0.5cm}
  \end{center}
  \begin{tabular}{@{}lllccccccccccc@{}}
    \hline
     \multicolumn{2}{c}{Peak Position} & & {Ident.} & \multicolumn{3}{c}{Integrated Fluxes} & & &\multicolumn{5}{c}{Dust Emissivity Exponent $\beta$} \\

RA     & Dec   & Source Name & Tracer & 1200    & 850      & 450     & note   & ref  & 10K & 20K & 30K & 40K & 50K\\(J2000)&(J2000)&   $^a$      &        &Jy$^{b}$ & Jy$^{b}$ & Jy$^{b}$& $^{c}$ &$^{d}$&     &     &     &     & \\ 
(1)&(2)&(3)&(4)&(5)&(6)&(7)&(8)&(9)&(10)&(11)&(12)&(13)&(14) \\
\hline
06 09 06.5 & +21 50 27 & G188.79+1.02 &  r & 1.9 & 8.4 & nocal &  & T  &  &  &  &  & 	\\
17 45 54.3 & -28 44 08 & G0.204+0.051 &  mm & 0.6 & 3.7 & 33.3 &  & PP & 3.5 & 2.8 & 2.6 & 2.5 & 2.4	\\
17 46 03.9 & -28 24 58 & G0.49+0.19 &  m  & 1.2 & 4.0 & 57.0 &  & W  & 3.1 & 2.4 & 2.2 & 2.1 & 2.1	\\
17 46 07.1 & -28 41 28 & G0.266-0.034 &  mm & 1.0 & 7.3 & 73.6 &  & PP & 3.8 & 3.1 & 2.9 & 2.8 & 2.7	\\
17 46 07.7 & -28 45 20 & G0.21-0.00 &  mr & 1.2 & 6.2 & 98.0 &  & W  & 3.8 & 3.1 & 2.9 & 2.8 & 2.7	\\
17 46 08.2 & -28 25 23 & G0.497+0.170 &  mm & 0.9 & 5.8 & 15.6 &  & H & 2.4 & 1.6 & 1.4 & 1.3 & 1.2	\\
17 46 09.5 & -28 43 36 & G0.240+0.008 &  mm & 7.2 & 25.5 & 216.9 &  & PP & 2.8 & 2.1 & 1.9 & 1.8 & 1.7	\\
17 46 10.1 & -28 23 31 & G0.527+0.181 &  r & 2.4 & 9.0 & 58.1 & $\diamond$ & H & 2.6 & 1.9 & 1.7 & 1.6 & 1.5	\\
17 46 10.7 & -28 41 36 & G0.271+0.022 &  mm & 0.6$^\alpha$ & 1.9 & 17.2 &  & PP & 2.7 & 2.0 & 1.8 & 1.7 & 1.6	\\
17 46 11.4 & -28 42 40 & G0.257+0.011 &  mm & 6.7 & 23.2 & 251.1 &  & PP & 2.9 & 2.2 & 2.0 & 1.9 & 1.9	\\
17 46 52.8 & -28 07 35 & G0.83+0.18 &  m  & 1.2 & 4.2 & 64.0 &  & W  & 3.2 & 2.5 & 2.3 & 2.2 & 2.2	\\
17 47 00.0 & -28 45 20 & G0.331-0.164 &  mm & 0.8 & 1.3 & nocal &$\diamond$ & H &  &  &  &  & 	\\
17 47 01.2 & -28 45 36 & G0.310-0.170 &  mm & 0.2 & 0.9 & 8.8 &  & PP & 3.2 & 2.5 & 2.3 & 2.2 & 2.1	\\
17 47 09.1 & -28 46 16 & G0.32-0.20 &  mr & 5.9 & 20.0 & 320.0 &  & W  & 3.2 & 2.5 & 2.3 & 2.2 & 2.2	\\
17 47 20.1 & -28 47 04 & G0.325-0.242 &  mm & 0.3 & 1.8 & ND&  & PP &  &  &  &  & 	\\
17 48 31.6 & -28 00 31 & G1.124-0.065 &  mm & 0.5 & 2.3 & ND&  & PP &  &  &  &  & 	\\
17 48 34.7 & -28 00 16 & G1.134-0.073 &  mm & 0.2 & 0.9 & ND&  & PP &  &  &  &  & 	\\
17 48 36.4 & -28 02 31 & G1.105-0.098 &  mm & 1.8 & 5.8 & 37.3 &  & PP & 2.4 & 1.7 & 1.5 & 1.4 & 1.3	\\
17 48 41.9 & -28 01 44 & G1.13-0.11 &  r & 7.9 & 13.4 & 174.3 &  & T  & 1.7 & 1.4 & 1.2 & 1.1 & 1.1	\\
17 48 48.5 & -28 01 13  & G1.14-0.12 &  m  & 0.2 & 1.9 & 31.0 &  & W  & 4.5 & 3.8 & 3.6 & 3.5 & 3.4	\\
17 50 14.5 & -28 54 31 & G0.55-0.85 &  mr & 15.8 & 52.0 & 850.0 &  & W  & 3.1 & 2.5 & 2.3 & 2.2 & 2.1	\\
17 50 18.8 & -28 53 14 & G0.549-0.868 &  mm & 0.4 & 2.3 & 15.4 & $\dagger$ & H & 3.1 & 2.4 & 2.2 & 2.1 & 2.0	\\
17 50 24.9 & -28 50 15 & G0.627-0.848 &  mm & 0.3 & 3.3 & 12.9 & $\star$ & H & 3.3 & 2.5 & 2.3 & 2.2 & 2.1	\\
17 50 26.7 & -28 52 23 & G0.600-0.871 &  mm & 0.8 & 3.4 & 20.9 & $\star$ & H & 2.7 & 2.0 & 1.8 & 1.7 & 1.6	\\
17 50 46.5 & -26 39 45 & G2.54+0.20 &  m  & 2.1 & 12.0 & 150.0 &  & W  & 3.7 & 3.0 & 2.8 & 2.7 & 2.6	\\
17 59 04.6 & -24 20 55 & G5.48-0.24 &  r & 1.1 & 7.4 & $<$9.7 & $\star$ & T  &  &  &  &  & 	\\
17 59 07.5 & -24 19 19  & G5.504-0.246 &  mm & 0.7 & 2.8 & 17.2 & $\diamond$ & H & 2.7 & 1.9 & 1.7 & 1.6 & 1.5	\\
18 00 31.0 & -24 03 59 & G5.89-0.39 &  r & 23.2 & 46.5 & 250.6 &  & T  & 1.6 & 1.0 & 0.7 & 0.6 & 0.6	\\
18 00 40.9 & -24 04 21 & G5.90-0.42 &  m  & 8.1 & 33.0 & 470.0 &  & W  & 3.4 & 2.7 & 2.5 & 2.4 & 2.3	\\
18 00 50.9 & -23 21 29 & G6.53-0.10 &  r & 3.0 & 11.1 & $<$7.8  &  & T  &  &  &  &  & 	\\
18 00 54.1 & -23 17 02 & G6.60-0.08 &  m  & 0.3 & 1.1 & 11.0 &  & W  & 3.0 & 2.3 & 2.0 & 2.0 & 1.9	\\
18 02 50.2 & -21 48 39 & $\ast$G8.111+0.257 &  mm & 0.2 & 0.6 & 6.3 & & H & 2.1 & 1.6	& 1.5 & 1.4 & 1.3	\\
18 02 52.8 & -21 47 54 & G8.127+0.255 &  mm & 0.9 & 1.9 & 15.7 & $\star$ & H & 1.9 & 1.3 & 1.1 & 1.0 & 0.9	\\
18 02 56.2 & -21 47 38 & G8.138+0.246 &  mm & 1.9 & 4.0 & 38.7 &  & H & 1.9 & 1.3 & 1.2 & 1.1 & 1.0	\\
18 03 00.8 & -21 48 10 & G8.13+0.22 &  mr & 8.0 & 19.0 & 240.0 & $\star$ & W  & 2.2 & 1.7 & 1.5 & 1.4 & 1.4	\\
18 03 26.3 & -24 22 29 & G5.948-1.125 &  mm & 0.3 & 1.0 & 7.2 &  & H & 2.6 & 1.8 & 1.6 & 1.5 & 1.5	\\
18 03 28.7 & -24 21 50 & $\ast$G5.962-1.128 & mm & 0.3 & 1.7	& 5.5 &  & H & 2.3 & 1.5 & 1.3 & 1.2	& 1.2	\\
18 03 34.5 & -24 21 41 & G5.975-1.146 &  mm & 0.5 & 1.3 & 11.5 & $\star$ & H & 2.3 & 1.7 & 1.5 & 1.4 & 1.3	\\
18 03 36.8 & -24 22 13 & G5.971-1.158 &  mm & 0.8 & 4.3 & 27.4 & $\star$ & H & 3.0 & 2.3 & 2.1 & 2.0 & 1.9	\\
18 03 40.9 & -24 22 37 & G5.97-1.17 &  r & 9.2 & 16.4 & 69.7 &  & T  & 1.3 & 0.6 & 0.4 & 0.3 & 0.3	\\
18 05 15.6 & -19 50 55 & G10.10+0.72 &  r & 0.6 & 1.4 & 13.0 &  & W  & 2.2 & 1.6 & 1.4 & 1.3 & 1.2	\\
18 06 14.8 & -20 31 37 & G9.63+0.19 &  mr & 8.5 & 30.0 & 400.0 &  & W  & 3.1 & 2.4 & 2.2 & 2.1 & 2.1	\\
18 06 18.9 & -21 37 21 & G8.68-0.36 &  mr & 13.9 & 42.0 & 730.0 &  & W  & 2.9 & 2.3 & 2.2 & 2.1 & 2.0	\\
18 06 23.5 & -21 36 57 & G8.686-0.366 &  m & 3.4 & 14.0 & 230.0 &  & W  & 3.5 & 2.8 & 2.6 & 2.5 & 2.5	\\
18 07 50.4 & -20 18 51 & G9.99-0.03 &  m  & 1.2 & 3.8 & 66.0 &  & W  & 3.1 & 2.5 & 2.3 & 2.2 & 2.1	\\
18 07 53.2 & -20 18 19 & G10.001-0.033 &  r & 0.4 & 0.9 & 17.0 &  & W  & 1.8 & 1.5 & 1.4 & 1.3 & 1.3	\\
18 08 37.9 & -19 51 41 & G10.47+0.02 &  mr & 24.0$^\alpha$  & 77.1 & 970.0 & $\diamond$ & W  & 2.9 & 2.3 & 2.1 & 2.0 & 1.9	\\
18 08 44.9 & -19 54 38 & G10.44-0.01 &  m  & 1.6 & 6.8 & 94.0 &  & W  & 3.4 & 2.7 & 2.5 & 2.4 & 2.4	\\
18 08 45.9 & -20 05 34 & G10.287-0.110 &  mm & 2.8 & 14.4 & 51.0 & $\diamond$ & H & 2.4 & 1.7 & 1.4 & 1.3 & 1.3	\\
18 08 49.4 & -20 05 58 & G10.284-0.126 &  m  & 2.6 & 9.8 & 53.9 &  & H & 2.5 & 1.7 & 1.5 & 1.4 & 1.4	\\
18 08 52.4 & -20 05 58 & G10.288-0.127 &  mm & 0.9 & 5.0 & 30.6 &  & H & 3.0 & 2.3 & 2.1 & 2.0 & 1.9	\\
18 08 55.5 & -20 05 58 & G10.29-0.14 &  mr & 7.8 & 29.0 & 148.2 &  & H & 2.4 & 1.7 & 1.4 & 1.3 & 1.3	\\
18 09 00.0 & -20 05 34 & G10.343-0.142 &  m  & 1.7 & 11.7 & edge &  & H &  &  &  &  & 	\\
18 09 03.5 & -20 02 54 & G10.359-0.149 &  mm & 1.4 & 4.6 & 52.2 &  & H & 2.9 & 2.2 & 2.0 & 1.9 & 1.9	\\
18 09 21.6 & -20 19 25 & G10.15-0.34 &  r & 6.5 & 10.6 & 83.1 &  & T  & 1.6 & 1.0 & 0.8 & 0.7 & 0.6	\\
18 10 15.7 & -19 54 45 & G10.63-0.33B &  mm & 1.4 & 4.7 & 45.0 &  & W  & 2.8 & 2.1 & 1.9 & 1.8 & 1.7	\\
18 10 18.0 & -19 54 05 & G10.62-0.33 &  m  & 3.7 & 12.7 & 67.0 &  & H & 2.3 & 1.6 & 1.4 & 1.3 & 1.2	\\
18 10 19.0 & -20 45 33 & G9.88-0.75 &  r & 5.5 & 21.9 & 195.5 & $\star$ & T  & 3.0 & 2.3 & 2.0 & 1.9 & 1.9	\\
18 10 24.1 & -20 43 09 & G9.924-0.749 &  mm & 0.7 & 1.7 & nocal &  & H &  &  &  &  & 	\\
18 10 29.4 & -19 55 41 & G10.62-0.38 &  mr & 27.9 & 76.0 & 1100.0 &  & W  & 2.6 & 2.0 & 1.9 & 1.8 & 1.7	\\
18 11 24.4 & -19 32 04 & G11.075-0.384 &  mm & 0.9 & 3.0 & nocal &  & H &  &  &  &  & 	\\
\end{tabular}
\end{table*}

\begin{table*}
\begin{center}
\contcaption{}
\vspace{-0.5cm}
  \end{center}
  \begin{tabular}{@{}lllccccccccccc@{}}
    \hline
     \multicolumn{2}{c}{Peak Position} & & {Ident.} & \multicolumn{3}{c}{Integrated Fluxes} & & &\multicolumn{5}{c}{Dust Emissivity Exponent $\beta$} \\

    RA & Dec              & Source Name & Tracer & 1200 & 850 & 450 & note& ref & 10K & 20K & 30K & 40K & 50K\\
(J2000)&(J2000)&   $^a$      &        &Jy$^{b}$ & Jy$^{b}$ & Jy$^{b}$& $^{c}$ &$^{d}$&     &     &     &     & \\  
(1)&(2)&(3)&(4)&(5)&(6)&(7)&(8)&(9)&(10)&(11)&(12)&(13)&(14) \\
\hline
18 11 31.8 & -19 30 44 & G11.11-0.34 &  r & 3.3 & 15.8 & 92.1 &  & T  & 2.8 & 2.0 & 1.8 & 1.7 & 1.6	\\
18 11 35.8 & -19 30 44 & G11.117-0.413 &  mm & 0.5 & 4.9 & $<$5.4 &  & T  &  &  &  &  & 	\\
18 11 51.4 & -17 31 30 & G12.88+0.48 &  m  & 6.9 & 19.0 & 280.0 &  & W  & 2.6 & 2.1 & 1.9 & 1.8 & 1.8	\\
18 11 53.6 & -17 30 02 & G12.914+0.493 &  mm & 0.7 & 1.9 & 13.7 &  & H & 2.3 & 1.6 & 1.4 & 1.3 & 1.2	\\
18 12 01.9 & -18 31 56 & G12.02-0.03 &  m  & 0.6 & 2.9 & 23.0 &  & W  & 3.1 & 2.4 & 2.2 & 2.1 & 2.0	\\
18 12 11.1 & -18 41 27 & G11.903-0.140 &  mr & 2.2 & 3.7 & 35.0 &  & H & 1.0 & 0.7 & 0.6 & 0.5 & 0.4	\\
18 12 15.6 & -18 44 58 & G11.861-0.183 &  mm & 0.1 & 0.6 & 6.5 & $\diamond$ & H & 3.7 & 2.9 & 2.7 & 2.6 & 2.6	\\
18 12 17.3 & -18 40 03 & G11.93-0.14 &  m  & 0.6 & 2.5 & 19.0 &  & W  & 2.9 & 2.2 & 1.9 & 1.8 & 1.8	\\
18 12 19.6 & -18 39 54 & G11.942-0.157 &  mm & 0.7 & 1.7 & 16.2 &  & H & 2.2 & 1.6 & 1.4 & 1.3 & 1.2	\\
18 12 39.2 & -18 24 17 & G12.20-0.09 &  mr & 4.3 & 17.0 & 270.0 & $\star$ & W  & 3.4 & 2.7 & 2.5 & 2.4 & 2.4	\\
18 12 42.7 & -18 25 08 & G12.18-0.12 &  m  & 0.6 & 7.2 & 22.5 & $\diamond$ & H & 3.2 & 2.4 & 2.2 & 2.1 & 2.0	\\
18 12 44.4 & -18 24 25 & G12.216-0.119 &  mm & 1.2 & 5.9 & 37.9 &  & H & 2.9 & 2.2 & 2.0 & 1.9 & 1.8	\\
18 12 51.2 & -18 40 40 & G11.99-0.27 &  m  & 0.3 & 1.7 & 22.0 &  & W  & 3.8 & 3.0 & 2.8 & 2.7 & 2.6	\\
18 12 56.4 & -18 11 04  & G12.43-0.05 &  r & 0.9 & 6.3 & $<$3.9 & $\star$ & T  &  &  &  &  & 	\\
18 13 54.7 & -18 01 41 & G12.68-0.18 &  m  & 5.6 & 14.6 & 115.0 & $\star$ & W  & 2.3 & 1.6 & 1.4 & 1.3 & 1.2	\\
18 13 58.5 & -18 54 21 & G11.94-0.62B &  mm & 4.5 & 14.0 & 66.1 &  & H & 2.1 & 1.4 & 1.2 & 1.1 & 1.0	\\
18 14 00.9 & -18 53 27 & G11.93-0.61 &  mr & 5.9 & 22.0 & 380.0 &  & W  & 3.4 & 2.7 & 2.5 & 2.4 & 2.4	\\
18 14 07.6 & -18 00 37 & G12.722-0.218 &  mm & 1.9 & 5.9 & 25.7 &  & H & 2.0 & 1.3 & 1.1 & 1.0 & 0.9	\\
18 14 34.3 & -17 51 56 & G12.90-0.25B &  mm & 1.6 & 3.8 & 53.3 &  & T  & 2.6 & 2.0 & 1.8 & 1.7 & 1.6	\\
18 14 36.1 & -17 54 56 & G12.859-0.272 &  mm & 2.3 & 5.4 & 47.7 &  & H & 2.1 & 1.5 & 1.3 & 1.2 & 1.2	\\
18 14 36.1 & -16 45 44 & G13.87+0.28 &  m & 6.0 & 16.7 & 100.1 &  & T  & 2.2 & 1.5 & 1.2 & 1.1 & 1.1	\\
18 14 39.5 & -17 52 00 & G12.90-0.26 &  m  & 8.6 & 25.0 & 300.0 &  & W  & 2.7 & 2.1 & 1.9 & 1.8 & 1.7	\\
18 16 22.1 & -19 41 27 & G11.49-1.48 &  m  & 4.9 & 19.0 & 210.0 &  & W  & 3.1 & 2.4 & 2.2 & 2.1 & 2.1	\\
18 17 00.5 & -16 14 44 & G14.60+0.01 &  mr & 2.2 & 9.6 & 176.0 & $\diamond$ & W  & 3.7 & 3.0 & 2.8 & 2.7 & 2.6	\\
18 19 12.6 & -20 47 31  & G10.84-2.59 &  r & 5.0 & 14.8 & 96.9 &  & T  & 2.4 & 1.6 & 1.4 & 1.3 & 1.2	\\
18 20 23.1 & -16 11 16 & G15.03-0.67 &  mr & 30.0 & 84.0 & 1880.0 & $\diamond$ & W  & 2.6 & 2.2 & 2.1 & 2.0 & 2.0	\\
18 21 14.6 & -14 32 52 & G16.580-0.079 &  mm & 0.5 & 1.8 & 3.3 & $\star$ & H & 1.4 & 0.6 & 0.4 & 0.3 & 0.2	\\
18 21 09.1 & -14 31 49 & G16.58-0.05 &  m  & 3.0 & 14.0 & 270.0 &  & W  & 3.8 & 3.1 & 2.9 & 2.8 & 2.8	\\
18 25 01.3 & -13 15 35 & G18.15-0.28 &  r & 2.5 & 4.1 & 118.7 &  & T  & 0.8 & 0.5 & 0.4 & 0.4 & 0.4	\\
18 25 07.3 & -13 14 23 & G18.177-0.296 &  mm & 0.9 & 2.8 & 29.7 &  & H & 2.8 & 2.1 & 1.9 & 1.8 & 1.7	\\
18 25 42.2 & -13 10 32 & G18.30-0.39 &  r & 5.6 & 13.1 & 255.1 & $\star$ & T  & 3.7 & 2.9 & 2.6 & 2.5 & 2.4	\\
18 27 16.3 & -11 53 51 & G19.61-0.13 &  m & 1.3 & 9.4 & 160.0 &  & W  & 4.3 & 3.6 & 3.4 & 3.3 & 3.2	\\
18 27 38.2 & -11 56 38 & G19.607-0.234 &  mr & 13.4 & 26.9 & 190.1 &  & H & 1.7 & 1.1 & 0.9 & 0.8 & 0.8	\\
18 27 55.5 & -11 52 39 & G19.70-0.27A &  m & 1.1 & 4.9 & 68.0 & $\star$ & W  & 3.5 & 2.8 & 2.6 & 2.5 & 2.4	\\
18 29 24.2 & -15 15 34 & G16.871-2.154 &  mm & -$^\alpha$ & 30.3 &164.7  &  & H &  &  &  &  & 	\\
18 29 24.4 & -15 16 04 & G16.86-2.15 &  m & 16.9$^\alpha$ & 58.2 & 266.4 &  & W  & 2.2 & 1.5 & 1.3 & 1.2 & 1.1	\\
18 29 33.1 & -15 15 50 & G16.883-2.188 &  mm & 0.5 & 1.7 & 12.4 & $\star$ & H & 2.6 & 1.9 & 1.7 & 1.6 & 1.5	\\
18 31 02.1 & -09 49 14 & G21.87+0.01 &  mr & 1.0 & 4.0 & 40.0 &  & W  & 3.1 & 2.4 & 2.2 & 2.1 & 2.0	\\
18 31 43.2 & -09 22 25 & G22.36+0.07B &  m & 2.5 & 4.5 & 51.0 & $\star$  & W  & 1.3  &0.9  &0.8  & 0.7 &0.7 \\
18 33 53.6 & -08 07 15  & G23.71+0.17 &  r & 3.0 & 15.1 & 56.2 &  & T  & 2.4 & 1.6 & 1.4 & 1.3 & 1.2	\\
18 33 53.6 & -08 08 43 & G23.689+0.159 &  mm & 0.4 & 1.4 & 12.9 & $\star$ & H & 2.8 & 2.1 & 1.9 & 1.8 & 1.8	\\
18 34 10.3 & -07 17 45  & G24.47+0.49 &  r & 3.8 & 21.8 & 292.1 & $\diamond$ & T  & 3.8 & 3.1 & 2.8 & 2.7 & 2.7	\\
18 34 20.9 & -05 59 40 & G25.65+1.04 &  mr & 6.5 & 25.0 & 220.0 &  & W  & 2.9 & 2.2 & 2.0 & 1.9 & 1.8	\\
18 34 31.3 & -08 42 47 & G23.25-0.24 &  m & 0.4 & 2.5 & 39.0 &  & W  & 4.1 & 3.3 & 3.1 & 3.0 & 3.0	\\
18 34 36.2 & -08 42 39 & G23.268-0.257 &  mm & 3.7 & 7.1 & 85.3 &  & H & 1.5 & 1.1 & 1.0 & 0.9 & 0.9	\\
18 34 39.4 & -08 31 33 & G23.43-0.18 &  m & 4.0 & 18.0 & 200.0 &  & W  & 3.3 & 2.6 & 2.4 & 2.3 & 2.2	\\
18 34 50.7 & -08 41 03 & G23.319-0.298 &  mm & 0.7 & 3.6 & ND & $\diamond$ & H &  &  &  &  & 	\\
18 36 06.1 & -07 13 47 & G23.754+0.095 &  mm & 1.1 & 3.0 & 38.5 &  & H & 2.6 & 2.0 & 1.8 & 1.7 & 1.7	\\
18 36 12.6 & -07 12 11 & G24.78+0.08 &  m & 13.6 & 34.0 & 400.0 &  & W  & 2.4 & 1.8 & 1.6 & 1.5 & 1.5	\\
18 36 18.4 & -07 08 52 & G24.84+0.08 &  m & 1.0 & 3.3 & 34.0 &  & W  & 2.8 & 2.1 & 1.9 & 1.8 & 1.8	\\
18 36 25.9 & -07 05 08 & G24.919+0.088 &  mm & 2.6 & 7.0 & 72.1 &  & H & 2.5 & 1.8 & 1.6 & 1.5 & 1.5	\\
18 38 03.0 & -06 24 01 & G25.70+0.04 &  mr & 1.9 & 5.3 & 103.0 & $\star$ & W  & 2.7 & 2.2 & 2.0 & 2.0 & 1.9	\\
18 39 03.6 & -06 24 10 & G25.82-0.17 &  m & 5.4 & 16.0 & 160.0 &  & W  & 2.6 & 2.0 & 1.8 & 1.7 & 1.6	\\
18 42 42.6 & -04 15 32 & G28.14-0.00 &  m & 0.8 & 4.6 & 35.0 &  & W  & 3.3 & 2.5 & 2.3 & 2.2 & 2.2	\\
18 42 54.9 & -04 07 40 & G28.287+0.010 &  mm & 0.5$^\alpha$ & 1.7 & 13.3 &  & H & 2.7 & 1.9 & 1.7 & 1.6 & 1.6	\\
18 42 58.1 & -04 13 56 & G28.20-0.04 &  mr & 7.0 & 25.0 & 300.0 &  & W  & 3.1 & 2.4 & 2.2 & 2.1 & 2.0	\\
18 43 02.4 & -04 14 59 & G29.193-0.073 &  mm & 0.3 & 1.3 & 8.9 &  & H & 2.8 & 2.1 & 1.9 & 1.8 & 1.7	\\
18 44 14.2 & -04  17 59 & G28.28-0.35 &  mr & 5.1 & 10.5 & 113.0 & $\star$ & W  & 1.8 & 1.3 & 1.1 & 1.1 & 1.0	\\
18 44 22.0 & -04 17 38 & G28.31-0.38 &  m & 1.1 & 4.0 & 34.0 & $\star$ & W  & 2.8 & 2.1 & 1.9 & 1.8 & 1.7	\\
18 45 52.8 & -02 42 29 & G29.888+0.001 &  mm & 1.0 & 2.9 & 14.2 & $\star$ & H & 2.1 & 1.3 & 1.1 & 1.0 & 1.0	\\
18 45 54.4 & -02 42 37 & G29.889-0.006 &  mm & 0.5 & 1.2 & 4.7 &  & H & 1.6 & 0.9 & 0.7 & 0.6 & 0.5	\\
18 45 59.7 & -02 41 17 & G29.918-0.014 &  mm & 0.3 & 1.2 & 10.1 & $\star$ & H & 2.9 & 2.2 & 2.0 & 1.9 & 1.9	\\
\end{tabular}
\end{table*}

\begin{table*}
\begin{center}
\contcaption{}
  \end{center}
\vspace{-0.5cm}
  \begin{tabular}{@{}lllccccccccccc@{}}
    \hline
     \multicolumn{2}{c}{Peak Position} & & {Ident.} & \multicolumn{3}{c}{Integrated Fluxes} & & &\multicolumn{5}{c}{Dust Emissivity Exponent $\beta$} \\

    RA & Dec              & Source Name & Tracer & 1200 & 850 & 450 & note& ref & 10K & 20K & 30K & 40K & 50K\\
(J2000)&(J2000)&   $^a$      &        &Jy$^{b}$ & Jy$^{b}$ & Jy$^{b}$& $^{c}$ &$^{d}$&     &     &     &     & \\ 
(1)&(2)&(3)&(4)&(5)&(6)&(7)&(8)&(9)&(10)&(11)&(12)&(13)&(14) \\
\hline
18 46 00.2 & -02 45 09 & G29.86-0.04 &  m & 1.0 & 5.0 & 30.9 &  & H & 2.9 & 2.2 & 2.0 & 1.9 & 1.8	\\
18 46 01.3 & -02 45 25 & G29.861-0.053 &  mm & 0.7 & 2.2 & 19.7 &  & H & 2.6 & 1.9 & 1.7 & 1.6 & 1.6	\\
18 46 02.4 & -02 45 57 & G29.853-0.062 &  mm & 0.8 & 3.2 & 25.6 &  & H & 2.9 & 2.2 & 2.0 & 1.9 & 1.8	\\
18 46 04.0 & -02 39 25 & G29.96-0.02B &  mr & 9.3 & 26.0 & 390.0 &  & W  & 2.7 & 2.1 & 1.9 & 1.9 & 1.8	\\
18 46 05.0 & -02 42 29 & G29.912-0.045 &  mm & 3.4 & 15.1 & 94.8 & $\diamond$& H & 2.8 & 2.1 & 1.9 & 1.8 & 1.7	\\
18 46 06.1 & -02 41 25 & G29.930-0.040 &  mm & 0.4 & 3.6 & 12.3 &  & H & 2.9 & 2.2 & 2.0 & 1.9 & 1.8	\\
18 46 09.8 & -02 41 25 & G29.937-0.054 &  mm & 1.2 & 2.1 & 11.8 &  & H & 1.3 & 0.7 & 0.5 & 0.4 & 0.4	\\
18 46 11.5 & -02 42 05 & G29.945-0.059 &  mm & 2.3 & 3.4 & 30.0 & $\star$ & H & 0.6 & 0.3 & 0.2 & 0.1 & 0.1	\\
18 46 12.5 & -02 39 09 & G29.978-0.050 &  m  & 1.9 & 9.3 & 59.0 &  & W  & 2.9 & 2.2 & 2.0 & 1.9 & 1.8	\\
18 46 58.6 & -02 07 27 & G30.533-0.023 &  mm & 0.6$^\gamma$ & 5.3 & 75.0 &  & T  & 4.3 & 3.6 & 3.4 & 3.3 & 3.2	\\
18 47 08.6 & -01 44 02 & G30.89+0.16 &  m  & 0.8 & 6.2 & 56.0 &  & W  & 3.8 & 3.0 & 2.8 & 2.7 & 2.7	\\
18 47 13.4 & -01 44 58 & G30.894+0.140 &  mm & 1.1 & 2.9 & 34.6 &  & H & 2.5 & 1.9 & 1.7 & 1.6 & 1.6	\\
18 47 18.9 & -02 06 07 & G30.59-0.04 &  m  & 3.2 & 12.0 & 113.0 &  & W  & 2.9 & 2.2 & 2.0 & 1.9 & 1.9	\\
18 47 26.7 & -01 44 42 & G30.924+0.092 &  mm & 0.5 & 1.1 & 12.0 &  & H & 2.1 & 1.5 & 1.3 & 1.3 & 1.2	\\
18 47 34.2 & -01 56 41 & G30.760-0.027 &  mm & 1.6 & 3.7 & ND &  & H &  &  &  &  & 	\\
18 47 34.3 & -01 12 47 & G31.41+0.30 &  mr & 15.2 & 40.0 & 113.0 &  & W  & 1.4 & 0.7 & 0.5 & 0.4 & 0.3	\\
18 47 35.4 & -02 02 07 & G30.682-0.072 &  mm & 2.5 & 5.5 & 76.6 &  & H & 1.9 & 1.5 & 1.4 & 1.3 & 1.2	\\
18 47 35.8 & -01 55 13 & G30.78-0.02 &  m  & 6.7 & 16.7 & 237.4 & $\star$ & T  & 2.7 & 2.0 & 1.8 & 1.7 & 1.7	\\
18 47 36.0 & -02 01 05 & G30.705-0.065 &  m & 8.4 & 29.0 & 460.0 &  & W  & 3.2 & 2.5 & 2.3 & 2.2 & 2.2	\\
18 47 37.9 & -01 57 45 & G30.76-0.05 &  mm & 2.3 & 9.1 & 120.0 & $\star$ & W  & 3.3 & 2.6 & 2.4 & 2.3 & 2.2	\\
18 47 39.2 & -01 58 41 & G30.740-0.060 &  mm & 3.4 & 8.9 & 110.5 &  & H & 2.5 & 1.9 & 1.7 & 1.6 & 1.6	\\
18 47 41.3  & -02 00 33 & G30.716-0.082 &  mm & 4.1 & 8.1 & 137.6 &  & H & 1.4 & 1.1 & 1.0 & 0.9 & 0.9	\\
18 47 41.3 & -01 59 45 & G30.729-0.078 &  mm & 1.4 & 4.4 & 42.5 &  & H & 2.7 & 2.0 & 1.8 & 1.7 & 1.7	\\
18 47 41.3 & -01 37 21 & G31.06+0.09 &  m  & 0.3 & 1.8 & $<$5  &  & W  &  &  &  &  & 	\\
18 47 46.5 & -01 54 16 & G30.81-0.05 &  m  & 16.9 & 54.0 & 610.0 &  & W  & 2.8 & 2.2 & 2.0 & 1.9 & 1.8	\\
18 48 01.6 & -01 36 01 & G31.119+0.029 &  mm & 0.2$^\alpha$ & 1.0 & 5.4 &  & H & 2.8 & 2.0 & 1.8 & 1.7 & 1.6	\\
18 48 12.4 & -01 26 23 & G31.28+0.06 &  mr & 5.3 & 15.0 & 170.0 &  & W  & 2.6 & 2.0 & 1.8 & 1.7 & 1.6	\\
18 48 09.7 & -01 27 50 & G31.256+0.061 &  mm & 0.8 & 4.0 & 19.9 & $\star$ & H & 2.7 & 2.0 & 1.7 & 1.6 & 1.6	\\
18 49 33.1 & -01 29 04  & G31.40-0.26 &  r & 3.1 & 10.4 & 160.4 &  & T  & 3.2 & 2.5 & 2.3 & 2.2 & 2.2	\\
18 49 34.2 & -01 29 44 & G31.388-0.266 &  mm & 0.1 & 0.7 & 4.9 &  & H & 3.3 & 2.6 & 2.3 & 2.2 & 2.2	\\
18 50 30.7 & -00 02 00 & G32.80+0.19 &  r & 9.2 & 23.8 & 314.0 &  & T  & 2.7 & 2.0 & 1.8 & 1.8 & 1.7	\\
18 52 08.0 & +00 08 10 & G33.13-0.09 &  mr & 2.7 & 13.7 & 117.9 &  & T  & 3.2 & 2.5 & 2.3 & 2.2 & 2.1	\\
18 52 50.2 & + 00 55 28 & G33.92+0.11 &  r & 6.2 & 21.0 & 226.0 &  & T  & 3.0 & 2.2 & 2.0 & 1.9 & 1.9	\\
18 53 18.2 & +01 14 57 & G34.256+0.155 &  m & 51.8 & 179.1 & 1655.6 & $\star$ & H & 2.8 & 2.1 & 1.9 & 1.8 & 1.8	\\
18 54 00.5 & +02 01 16 & G35.02+0.35 &  mr & 2.5 & 10.0 & 128.3 &  & T  & 3.3 & 2.6 & 2.4 & 2.3 & 2.2	\\
18 56 01.2 & +02 22 59 & G35.57+0.07 &  r & 1.5 & 4.3 & 43.8 &  & H & 2.6 & 1.9 & 1.7 & 1.6 & 1.6	\\
18 56 03.9 & +02 23 23 & G35.586+0.061 &  mm & 0.5 & 1.8 & 16.8 & $\star$ & H & 2.9 & 2.2 & 2.0 & 1.9 & 1.8	\\
18 56 05.5 & +02 22 27 & G35.575+0.048 &  mm & 0.3 & 1.2 & 7.8 &  & H & 2.7 & 2.0 & 1.8 & 1.7 & 1.6	\\
18 57 09.0 & +01 39 05 & G35.05-0.52 &  r & 0.7 & 5.8 & nocal & $\star$ & T  &  &  &  &  & 	\\
19 00 06.9 & +03 59 39 & G37.475-0.106 &  m  & 0.2 & 0.6 & 6.1 &  & H & 2.6 & 2.0 & 1.8 & 1.7 & 1.6	\\
19 23 32.6 & +14 29 52 & G49.456-0.354 &  mm & 3.6 & 11.7 & 83.8 &  & H & 2.5 & 1.8 & 1.6 & 1.5 & 1.4	\\
19 23 35.9 & +14 31 04  & G49.482-0.355 &  mm & 1.2 & 5.1 & 40.3 &  & H & 3.0 & 2.2 & 2.0 & 1.9 & 1.9	\\
19 23 38.2 & +14 34 16 & G49.533-0.338 &  mm & 0.2 & 0.8 & 5.6 &  & H & 2.8 & 2.0 & 1.8 & 1.7 & 1.7	\\
19 23 43.1 & +14 30 32 & G49.488-0.385 &  mm & 71.5 & 132.8 & 1154.0 &  & H & 1.5 & 1.0 & 0.8 & 0.8 & 0.7	\\
19 23 45.9 & +14 29 44 & G49.481-0.401 &  mm & 6.2 & 13.9 & ND  &  & H &  &  &  &  & 	\\
19 23 49.2 & +14 28 48 & G49.474-0.420 &  mm & 0.9 & 4.9 & 50.3 &  & H & 3.5 & 2.8 & 2.6 & 2.5 & 2.4	\\
19 23 50.8 & +14 29 52 & G49.494-0.420 &  mm & 0.6 & 1.5 & 17.7 &  & H & 2.4 & 1.8 & 1.6 & 1.5 & 1.5	\\
19 43 10.0 & +23 44 59 & G59.794+0.076 &  mm & 0.8 & 2.4 & 21.1 &  & H & 2.6 & 1.9 & 1.7 & 1.6 & 1.5	\\
19 43 11.2 & +23 44 03 & G59.78+0.06 &  r & 4.7 & 21.0 & 132.5 &  & H &2.8  & 2.1 & 1.9 & 1.8 & 1.7	\\
\hline
\end{tabular}

\scriptsize
\begin{flushleft}

$^a$ Source names are consistent with column 3 of Table 5 in Paper I. Source names given to two (or fewer) decimal places are consistent with those reported by \citeauthor{walsh98, minier01, thompson06} from which they were targeted. Source names given to three decimal places denote sources identified in the SIMBA survey (Paper I), with the extended Galactic name intended to distinguish closely associated sources. Source names preceded with an asterisk ($\ast$) are those sources which were not reported in paper I. 

$^b$ Integrated fluxes in Jy. $^{\alpha}$denotes sources that have more than one millimetre peak encompassed, for which it was not possible to clearly distinguish the individual cores. The flux quoted here is for all the sources. $^{\zeta}$denotes sources located too close to the edge of the map, for their fluxes to be calculated. In the majority of cases it is not possible to determine the peak of the millimetre emission either. Note: this footnote was originally $\beta$ in Paper I, but has been reassigned here in order to avoid confusion with the dust grain emissivity index. $^{\gamma}$denotes sources situated quite close to the edge of the map, with some uncertainty in source size. The fluxes quoted here are a lower limit. Sources with 850\microm\, emission but without 450\microm emission are denoted in the table, with `no cal' indicating no suitable calibrator for the data, `ND' indicating no detection, and upper limits included for the remainder of the sources.

$^c$ Designations in this column indicate that the SIMBA core has been resolved into multiple components by SCUBA. A $\star$, $\diamond$, $\dagger$ indicates the presence of 2, 3 and 4 SCUBA cores respectively.

$^d$ Indicates the origin of the SCUBA data per corresponding SIMBA core. A `T' in this column indicates the survey of \citet{thompson06}, `PP' indicates the Galactic Plane map of \citet{p-p00}, `W' indicates the SCUBA data of \citet{walsh03}, and `H' indicates sources detected in this survey (i.e. Hill et al.) as reported in Table \ref{chap:scuba:scubafluxes}.\\
\end{flushleft}
\end{table*}
\normalsize

\section[]{Discussion}\label{chap:scuba:discussion}
\subsection[]{Analysis of the results}\label{chap:scuba:analysis}

   The aim of this SCUBA continuum survey was to obtain the complementary submillimetre data for the 405 sources in the SIMBA source list, in order to determine the dust grain emissivity exponent $\beta$, and to facilitate future spectral energy distribution (SED) analysis (to be presented in Paper III). The SIMBA source list is comprised of four distinct classes of source, as described in Section \ref{chap:scuba:simscuba}.

   As described in Section \ref{chap:scuba:results:scuba} and Table \ref{chap:scuba:scubafindings}, also detected in this survey are 106 submillimetre sources, which are devoid of methanol maser and radio continuum sources. 50 per cent of these submillimetre sources are a component of an existing SIMBA core from Paper I, that is, the higher resolution SCUBA instrument has resolved a SIMBA core into multiple submillimetre components. The remaining 53 submillimetre sources are also devoid of SIMBA millimetre emission, and hence are detected solely from their submillimetre continuum emission in this survey (see Section \ref{chap:scuba:results:scuba}). Analysis of these sources is explored further in Section \ref{chap:scuba:discussion:beta:submm}.
 
   This survey, together with existing SCUBA data \citep{p-p00, walsh03, thompson06}, reasonably samples the SIMBA source list of 405 sources. 28 per cent of the source list is below the declination limit of the JCMT, thus these sources are currently unobtainable at submillimetre wavelengths. Of the remaining 292 sources in Table \ref{chap:scuba:tab_simscuba}, 154 (53 per cent) sources have submillimetre data at both 450 and 850\microm, to complement that of SIMBA (1.2\,mm), while 17 others have complementary 850\microm\, data only.

   The dust grain emissivity index $\beta$ was determined using a Levenberg-Marquardt least squares fit, as discussed in Section \ref{chap:scuba:beta_exp}, for a range of temperatures characteristic of cold cores (10\,-\,50\,K). The results of this fitting, and hence the value of $\beta$ for each of these temperatures, is reported in columns 9 through 14 of Table \ref{chap:scuba:tab_simscuba}.

   The robustness of the fits is given by the $\chi^2$ statistic. For a Gaussian distribution of two free parameters, $\beta$ and the constant `A', the appropriate value of $\Delta \chi^2$ should be less than 2.3 \citep*{lampton76}. This value gives a confidence of 0.68 (or 1-$\sigma$) that the correct values of the fitted parameters will be included in the set of acceptable solutions. For the temperature range 20\,-\,50\,K, 85 per cent of the sample is less than this $\Delta \chi^2$ value of 2.3, whilst 95 per cent is included in the 0.90 confidence level (1.6-$\sigma$) value of 4.61 \citep{lampton76}. At 10\,K, 80 per cent of the sample is within the 1-$\sigma$ confidence interval, whilst 94 per cent of the sample is within the 1.6-$\sigma$ confidence level. 

   The error estimate in the value of the dust grain emissivity exponent $\beta$ for each of the sources reported in Table \ref{chap:scuba:tab_simscuba} is approximately 20 per cent at 10\,K and 30 per cent for the 20\,-\,50\,K range. The mean, median and standard deviation values for each temperature for the sample, as well as for each class of source, are reported in Table \ref{chap:scuba:averages}. At 20\,K, the mean value of $\beta$ for the sample is 2.0, with a standard deviation of 0.6.

\begin{table}
\begin{center}
\caption{Mean, median and standard deviations of $\beta$ for each of the temperatures analysed (10\,-\,50\,K). \label{chap:scuba:averages}}
\end{center}
\vspace{-0.5cm}
\begin{tabular} {@{}llccccc@{}}
\hline
Parameter & Source Class & 10K & 20K & 30K & 40K & 50K\\
\hline
Mean  & Sample & 2.7 & 2.0 & 1.8 & 1.8 & 1.7\\
& mm-only & 2.6 & 1.9 & 1.7 & 1.6 & 1.6 \\
& maser &3.1 & 2.4 & 2.2 & 2.1 & 2.0 \\
& maser+radio& 2.8 & 2.1 & 1.9 & 1.9 & 1.8 \\
& radio & 2.4 & 1.8 & 1.6 & 1.5 & 1.4\\
\hline
Median & Sample &2.8 & 2.1 & 1.9 & 1.8 & 1.7 \\
& mm-only & 2.7 & 2.0 & 1.8 & 1.7 & 1.6\\
& maser & 3.0 & 2.3 & 2.1 & 2.0 & 2.0 \\
& maser+radio & 2.9 & 2.3 & 2.1 & 2.0 & 2.0\\
& radio& 2.6 & 1.9 & 1.7 & 1.6 & 1.5\\
\hline
Standard & Sample & 0.7 & 0.6 & 0.6 & 0.6 & 0.6\\
Deviation & mm-only & 0.6 & 0.6 & 0.6 & 0.5 & 0.5\\
& maser & 0.6 & 0.5 & 0.5 & 0.5 & 0.5\\
& maser+radio & 0.6 & 0.6 & 0.6 & 0.6 & 0.6\\
& radio & 0.8 & 0.7 & 0.7 & 0.7 & 0.7\\
\hline
\end{tabular}
\end{table}

\begin{figure*}
  \begin{minipage}{1.0\textwidth}
         \includegraphics[width=7.0cm, height=7.0cm]{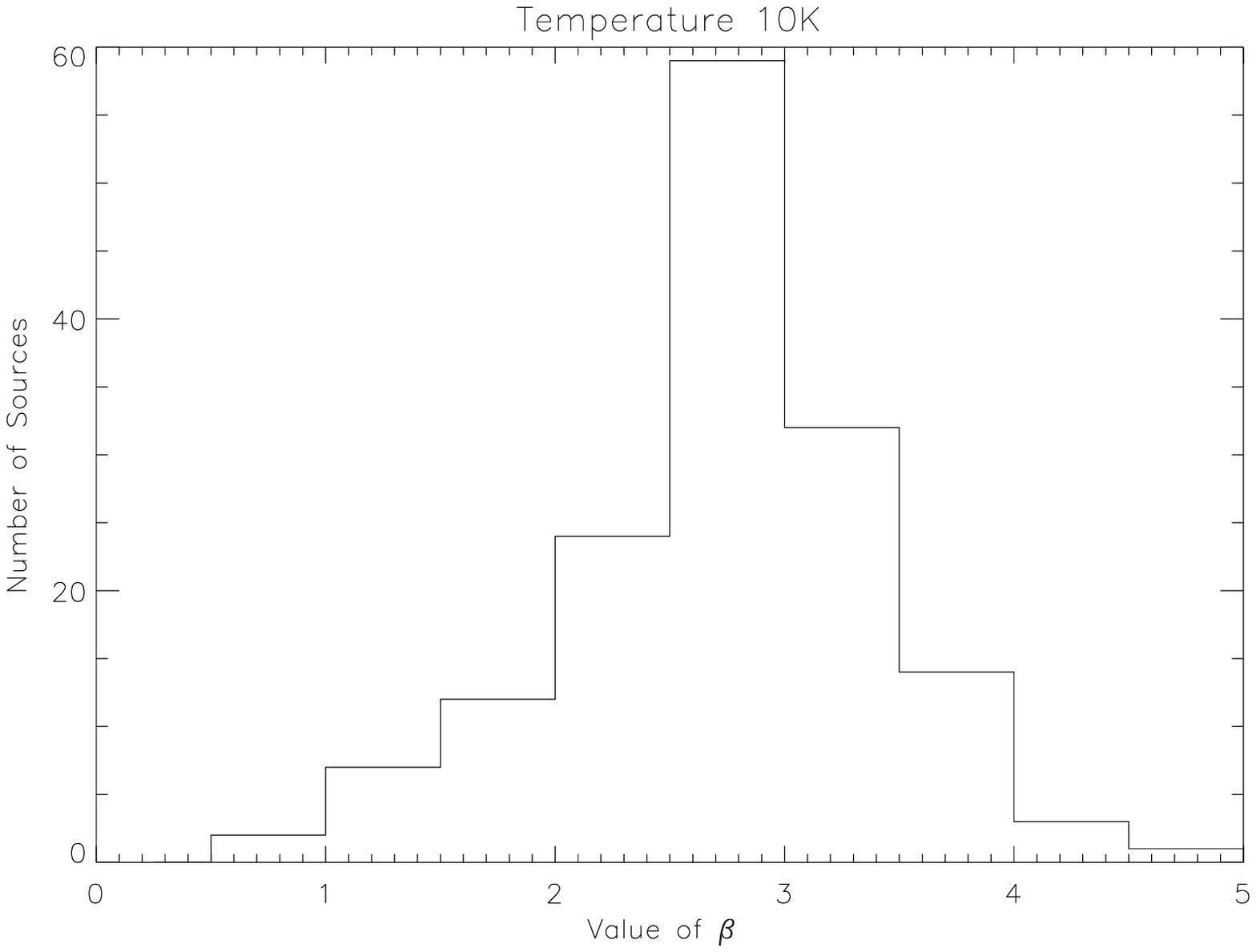}
	 \includegraphics[width=7.0cm, height=7.0cm]{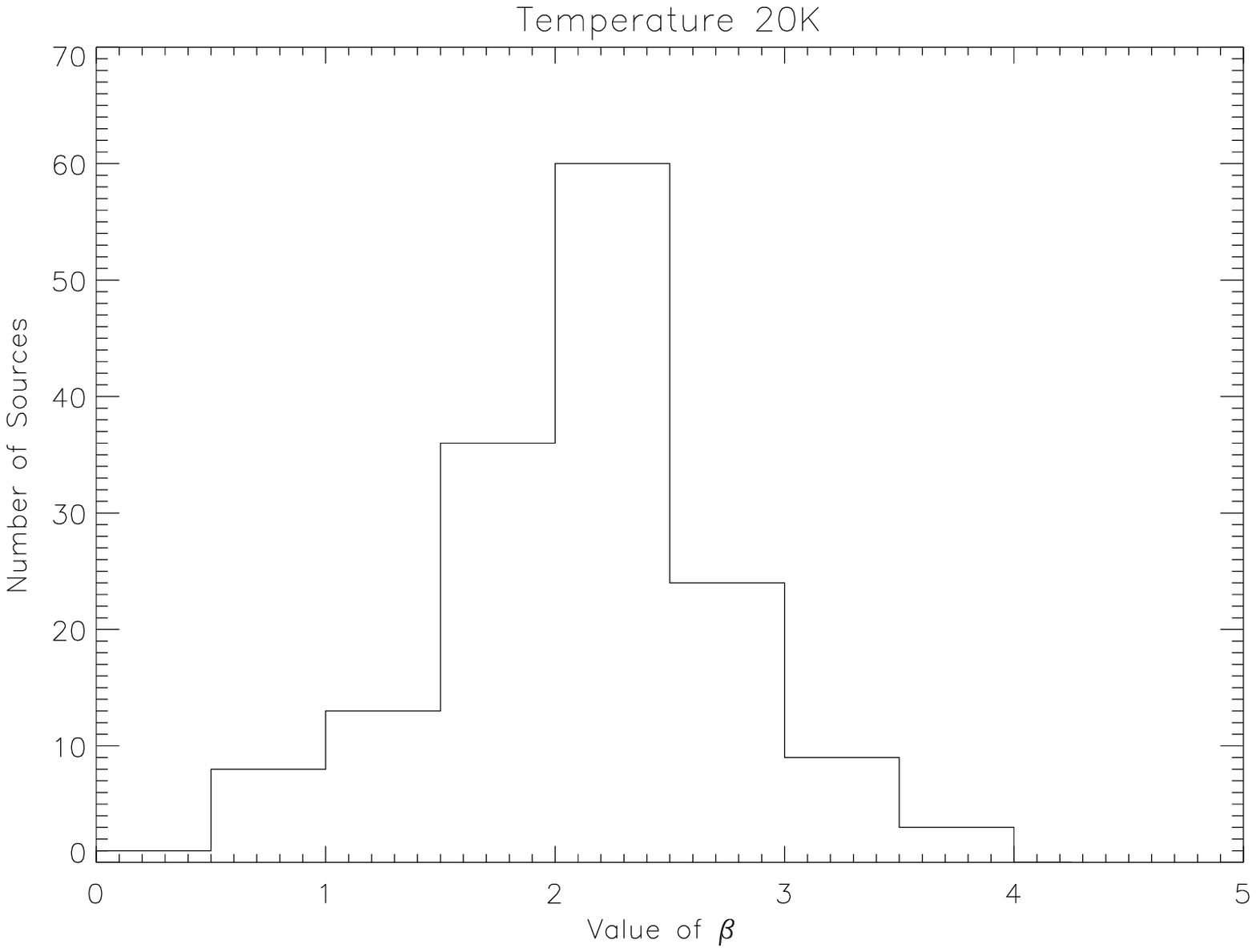}
  \end{minipage}
  \begin{minipage}{1.0\textwidth}
         \includegraphics[width=7.0cm, height=7.0cm]{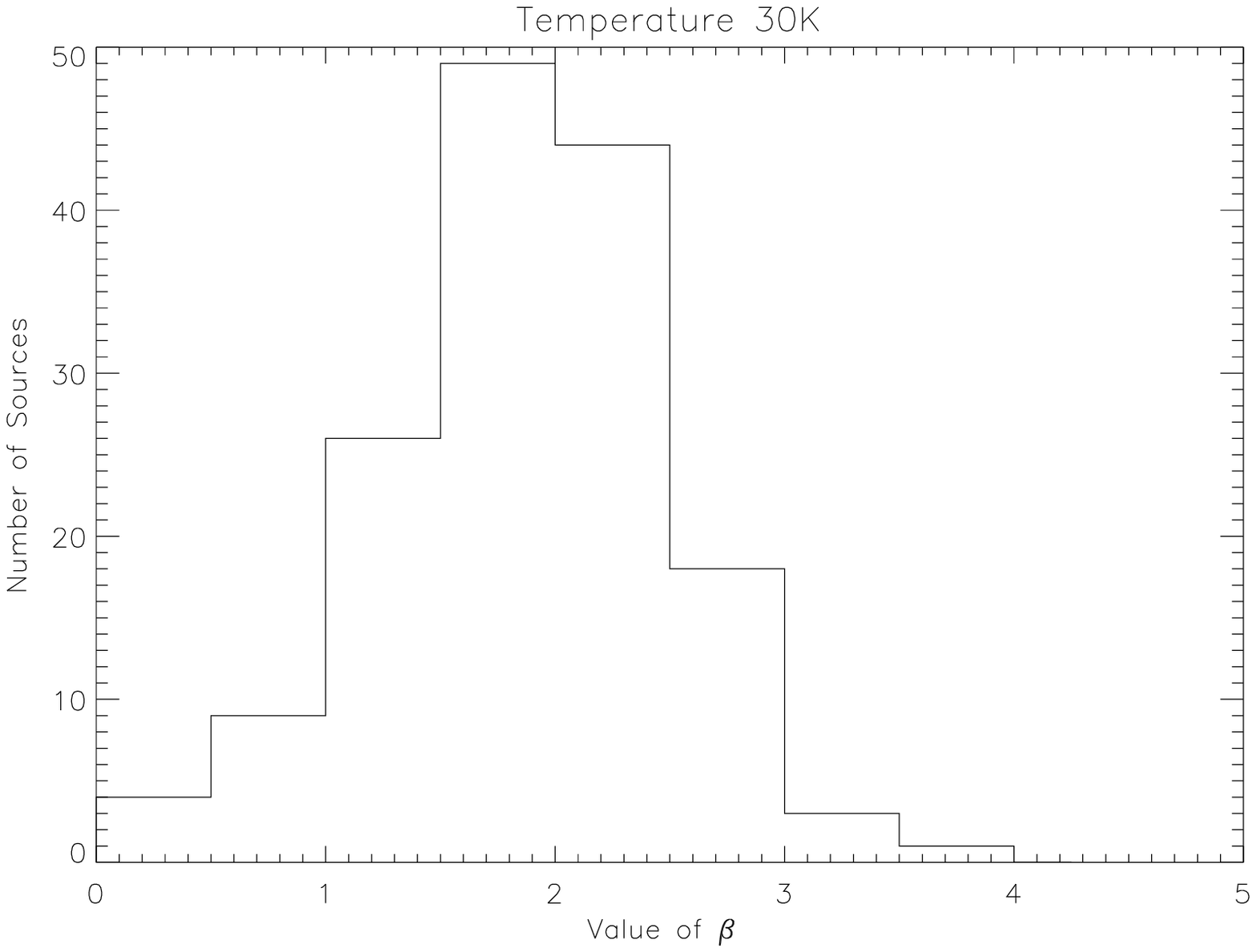}
	 \includegraphics[width=7.0cm, height=7.0cm]{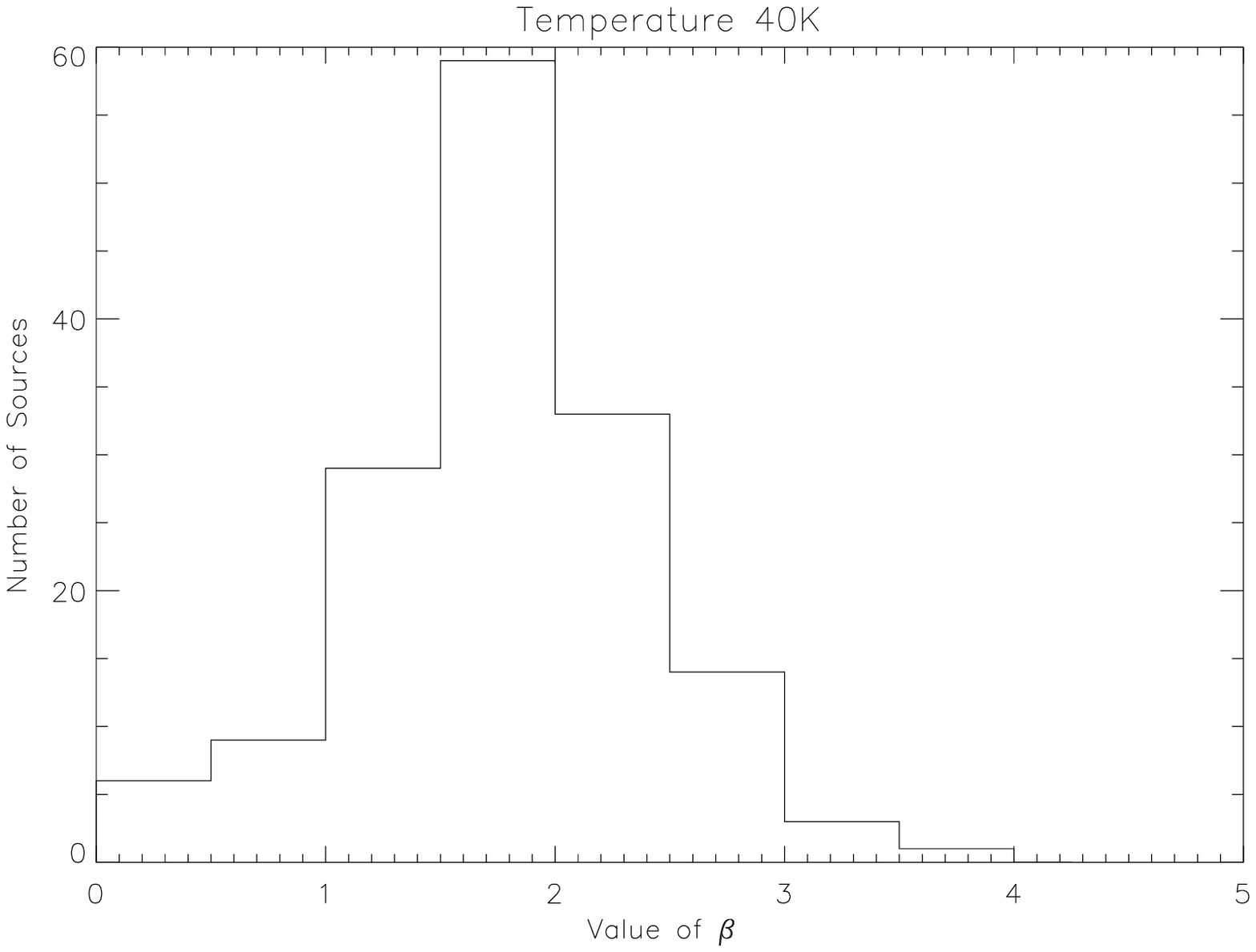}
  \end{minipage}
  \begin{minipage}{1.0\textwidth}
         \includegraphics[width=7.0cm, height=7.0cm]{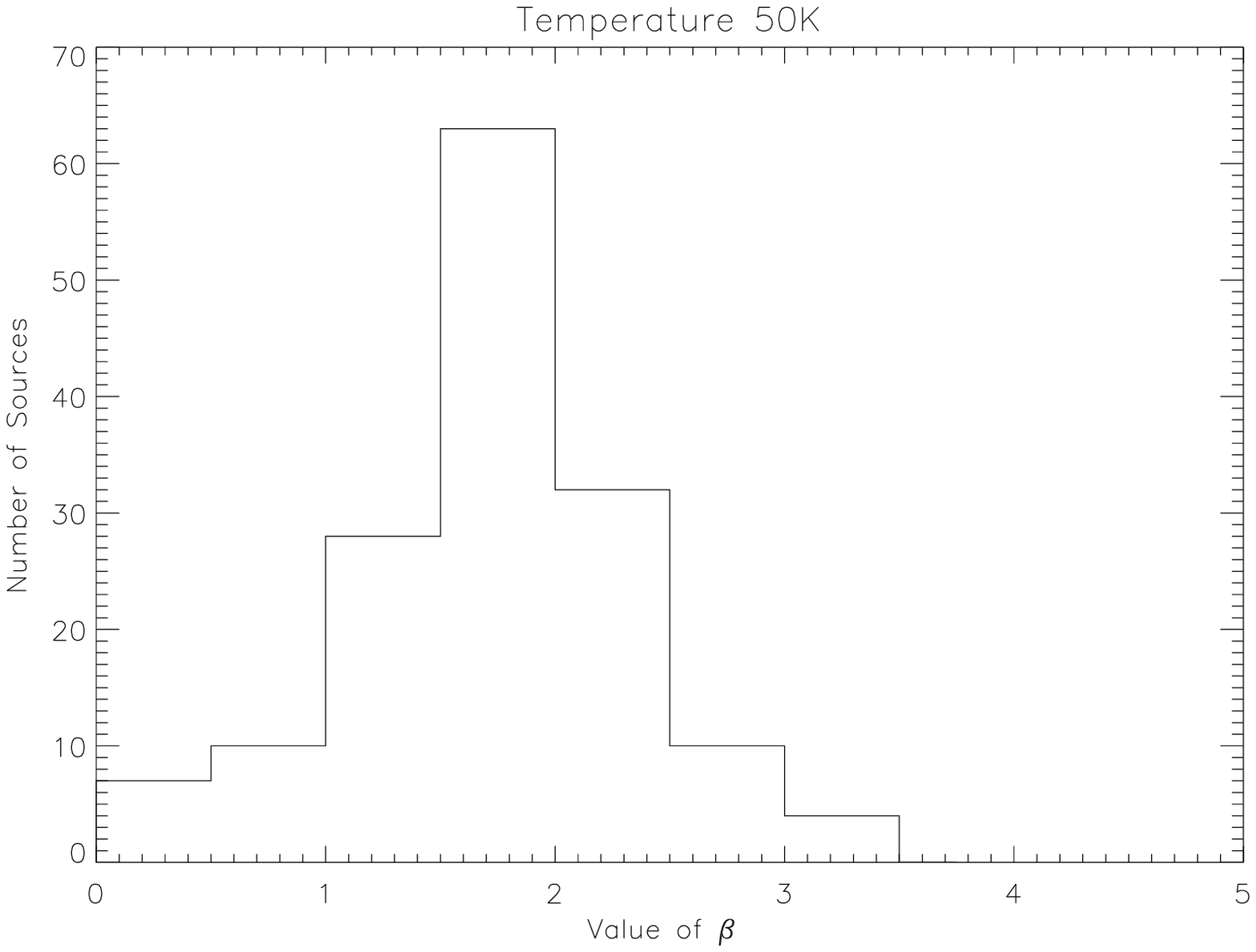}
  \end{minipage}
  \vspace{+0.1cm}
  \caption{Histograms of the dust grain emissivity index $\beta$ with respect to temperature. The temperature is indicated on each plot. Typical standard deviation for each plot is about 0.6 as determined for the sample.}
\label{chap:scuba:hist_temperature}
\end{figure*}

   The histogram of $\beta$ with respect to the temperature of the core is found in Figure \ref{chap:scuba:hist_temperature}. From these plots, it can be seen that the mean value of $\beta$ is dependent on the temperature assumed when fitting the data. An increase in assumed temperature (from 10\,-50\,K) results in a decrease of the mean value of $\beta$. The standard deviation of $\beta$ at each temperature (see Table \ref{chap:scuba:averages}) is typically of order of the distribution of $\beta$ for each temperature. Therefore, the error in the sample for $\beta$ accounts for the variation of $\beta$ with respect to the temperature.

\begin{figure*}
 \begin{minipage}{1.0\textwidth}
         \includegraphics[width=7.0cm, height=7.0cm]{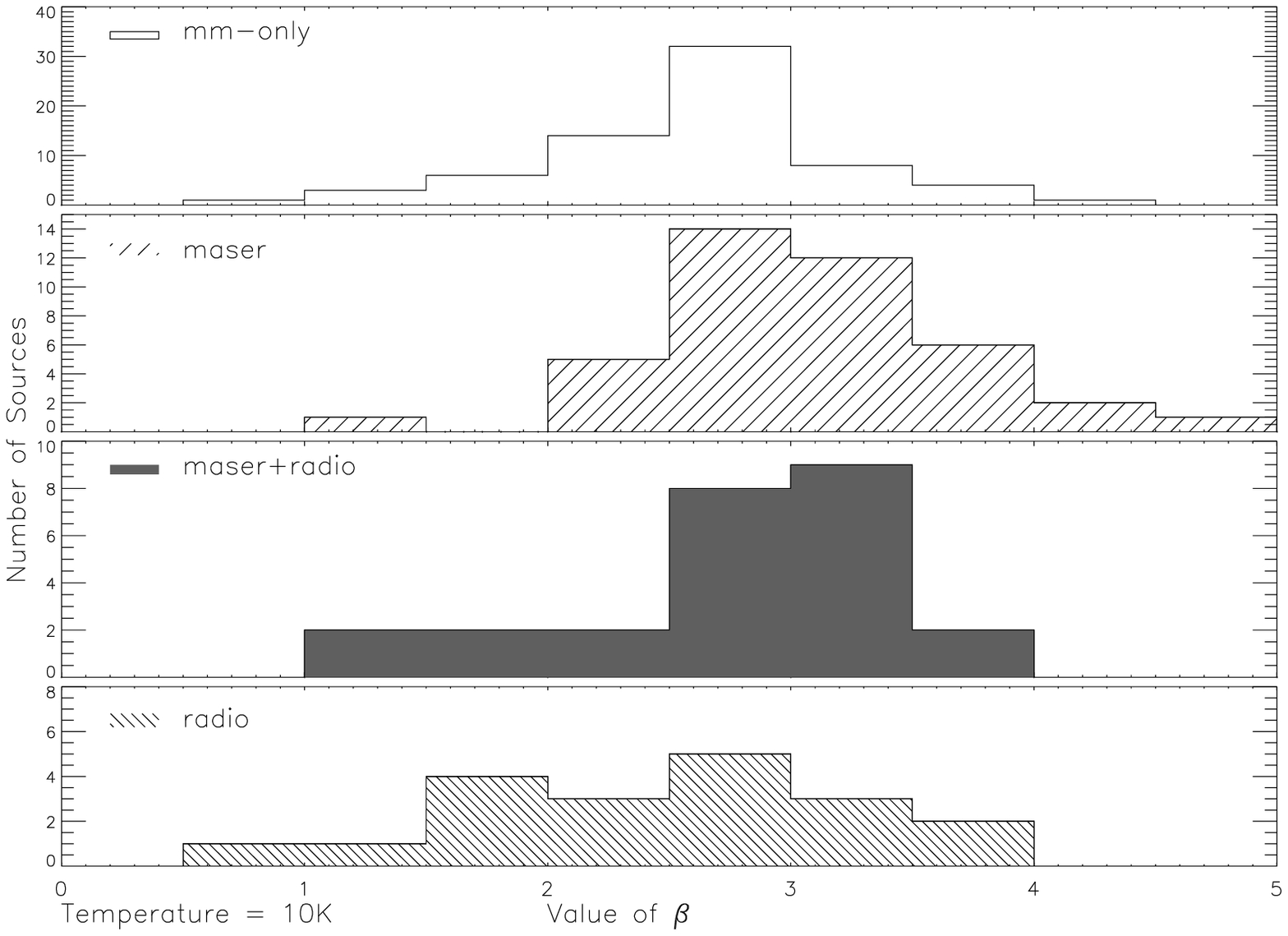}
	 \includegraphics[width=7.0cm, height=7.0cm]{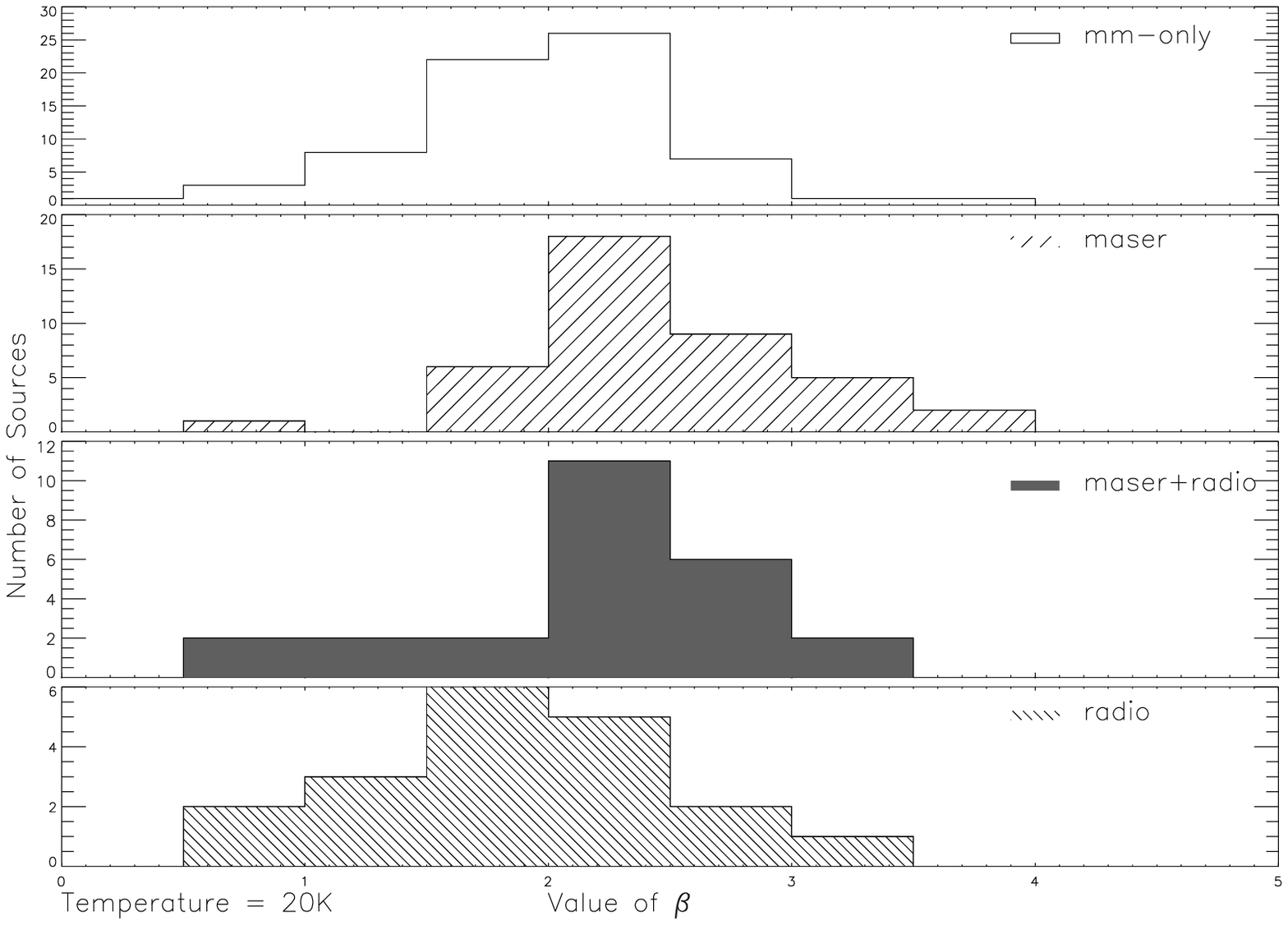}
  \end{minipage}
  \begin{minipage}{1.0\textwidth}
         \includegraphics[width=7.0cm, height=7.0cm]{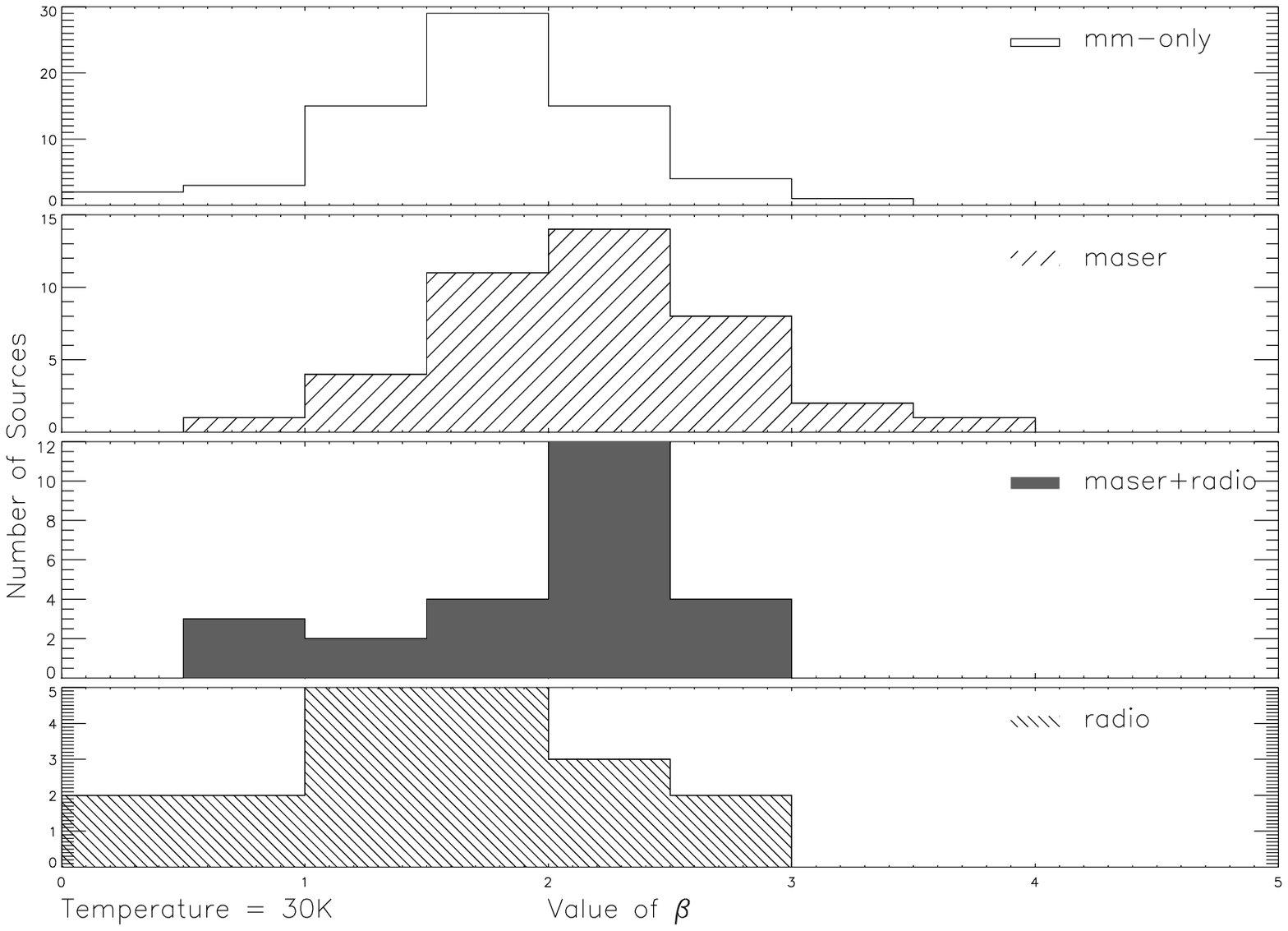}
	 \includegraphics[width=7.0cm, height=7.0cm]{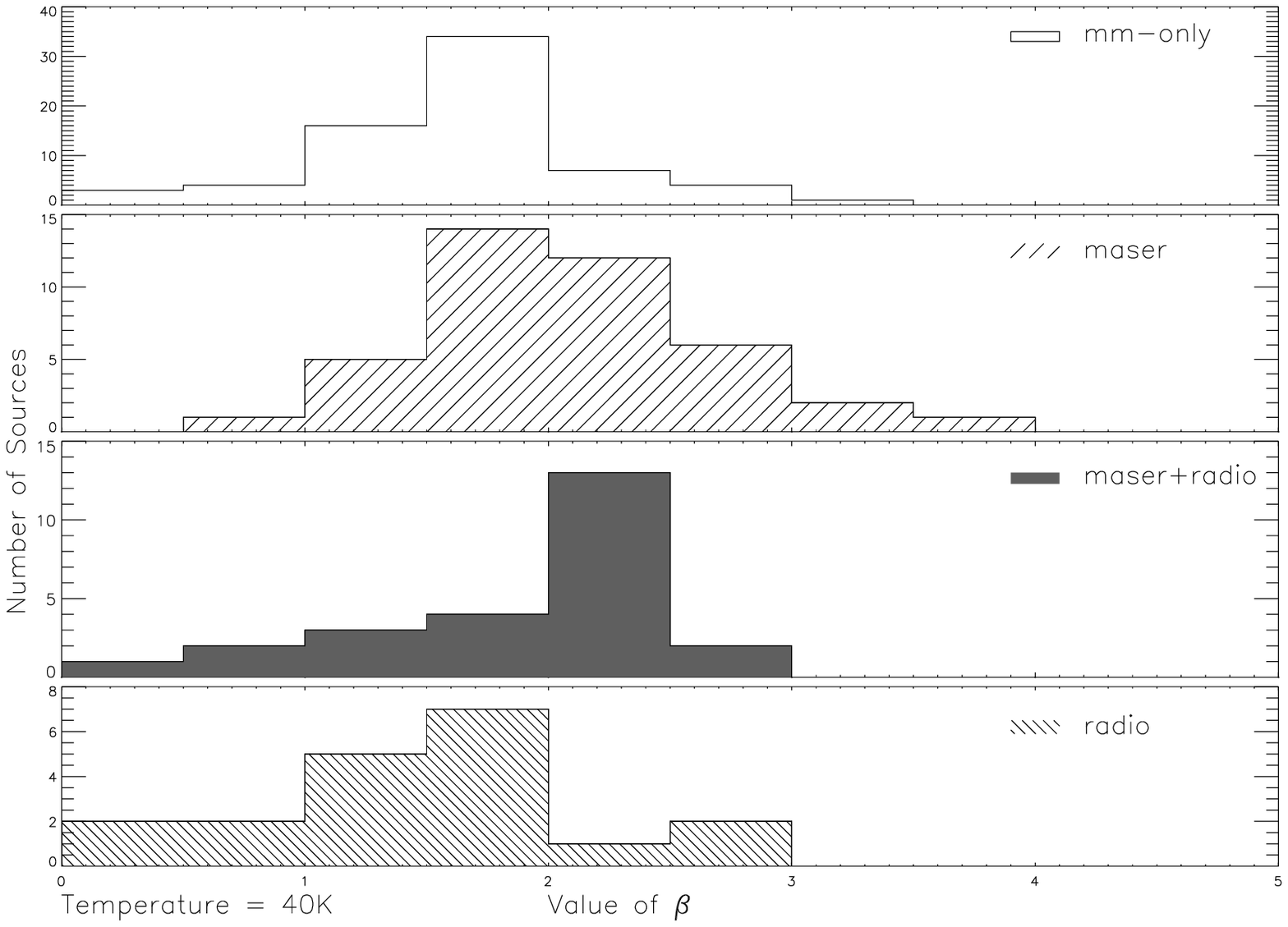}
  \end{minipage}
  \begin{minipage}{1.0\textwidth}
         \includegraphics[width=7.0cm, height=7.0cm]{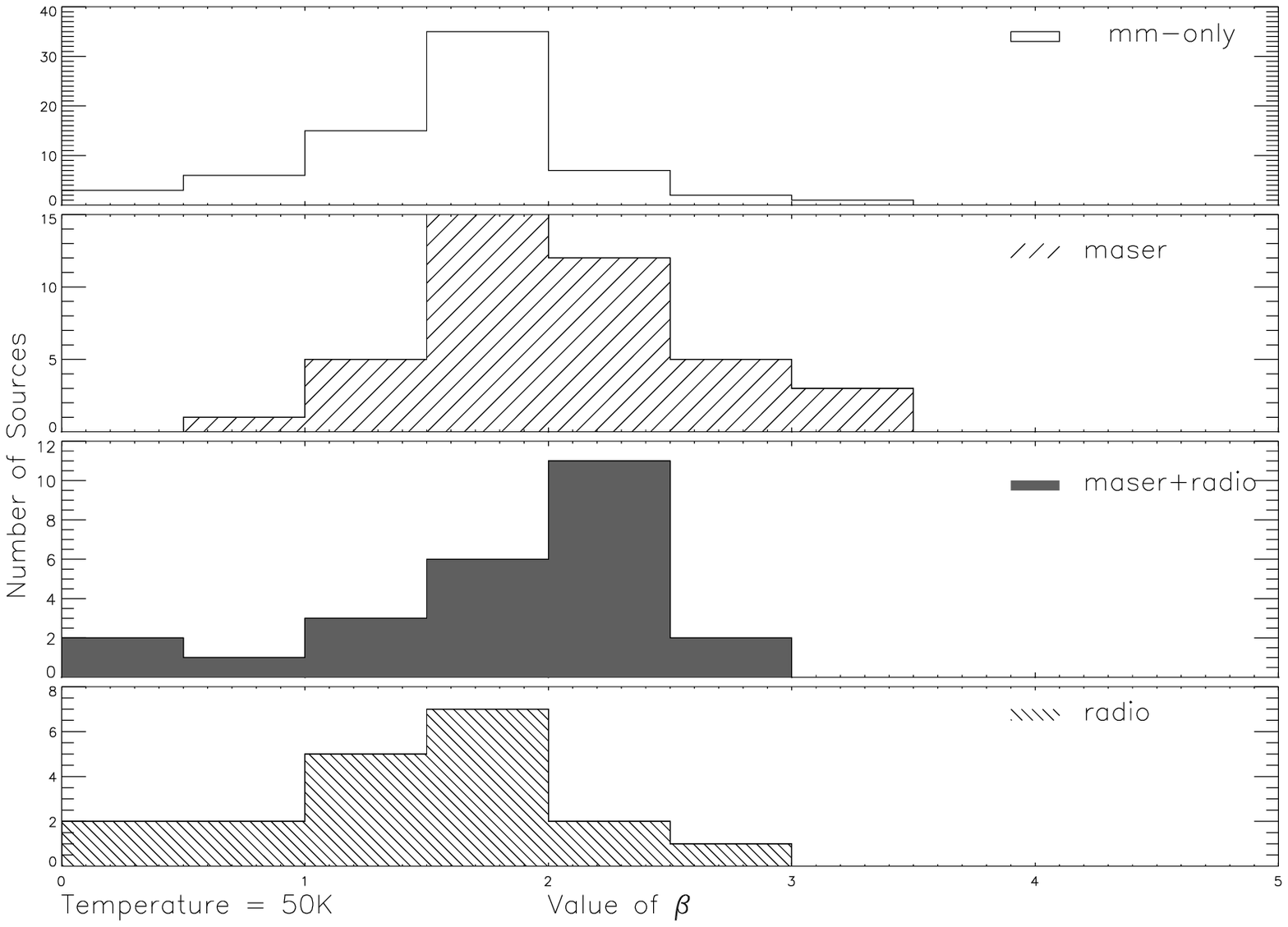}
  \end{minipage}
  \vspace{+0.1cm}
  \caption{Histograms of the dust grain emissivity index with respect to the star formation tracer associated with each SIMBA core. The different classes of source are as depicted on the plots, with the temperature indicated in the lower left corner of each plot.}
\label{chap:scuba:hist_tracer}
\end{figure*}

   The histogram of the dust grain emissivity index $\beta$ with respect to the type of core in the sample is presented in Figure \ref{chap:scuba:hist_tracer}. It can be seen from Fig. \ref{chap:scuba:hist_tracer} and from Table \ref{chap:scuba:averages} that $\beta$ varies little across the four classes of source in the sample  (mm-only, maser, maser+radio, radio) for each temperature. There is little difference in the mean/median value of $\beta$ across temperature.

  However, it can been seen in Fig. \ref{chap:scuba:hist_tracer} that the maser component of the sample has a `tail-end' population of higher values of $\beta$, that is not present in the other classes of source.

   Examination of the sources in this `tail-end' maser population, which were all observed in the SCUBA survey of \citet{walsh03}, reveals them to have 450/850\microm\, flux ratios typical of the rest of the sample reported by \citet{walsh03}. The particular SCUBA maps harbouring these maser sources are not typically noisier on average than the maps of the other maser sources reported in the \citet{walsh03} survey.  

   Comparison of the 450 and 850\microm\, fluxes of these `tail-end' maser sources with the 1.2\,mm flux reported in Paper I, reveals these particular maser sources to have much higher 850/1200\microm\, and 450/1200\microm\, flux ratios than other sources in the SIMBA survey. Typically we would expect the 850\microm\, to be 3 times brighter than SIMBA and the 450\microm\, to be 30 times brighter (see Section \ref{chap:scuba:results:scuba}). These particular `tail-end' maser sources are on average 7\,-\,10 times brighter at 850\microm\, and  $\ge$~90 times brighter at 450\microm\, compared to the SIMBA 1.2\,mm flux. The difference in flux ratios most likely arises from an error in the calibration factors for one or both of the SCUBA and SIMBA data for these sources. We therefore, do not believe this `tail-end' maser population to be significant. This could arise, for instance, if light cloud was experienced for some of the 1.2\,mm measurements.

\begin{figure*}
 \begin{minipage}{1.0\textwidth}
         \includegraphics[width=7.0cm, height=7.0cm]{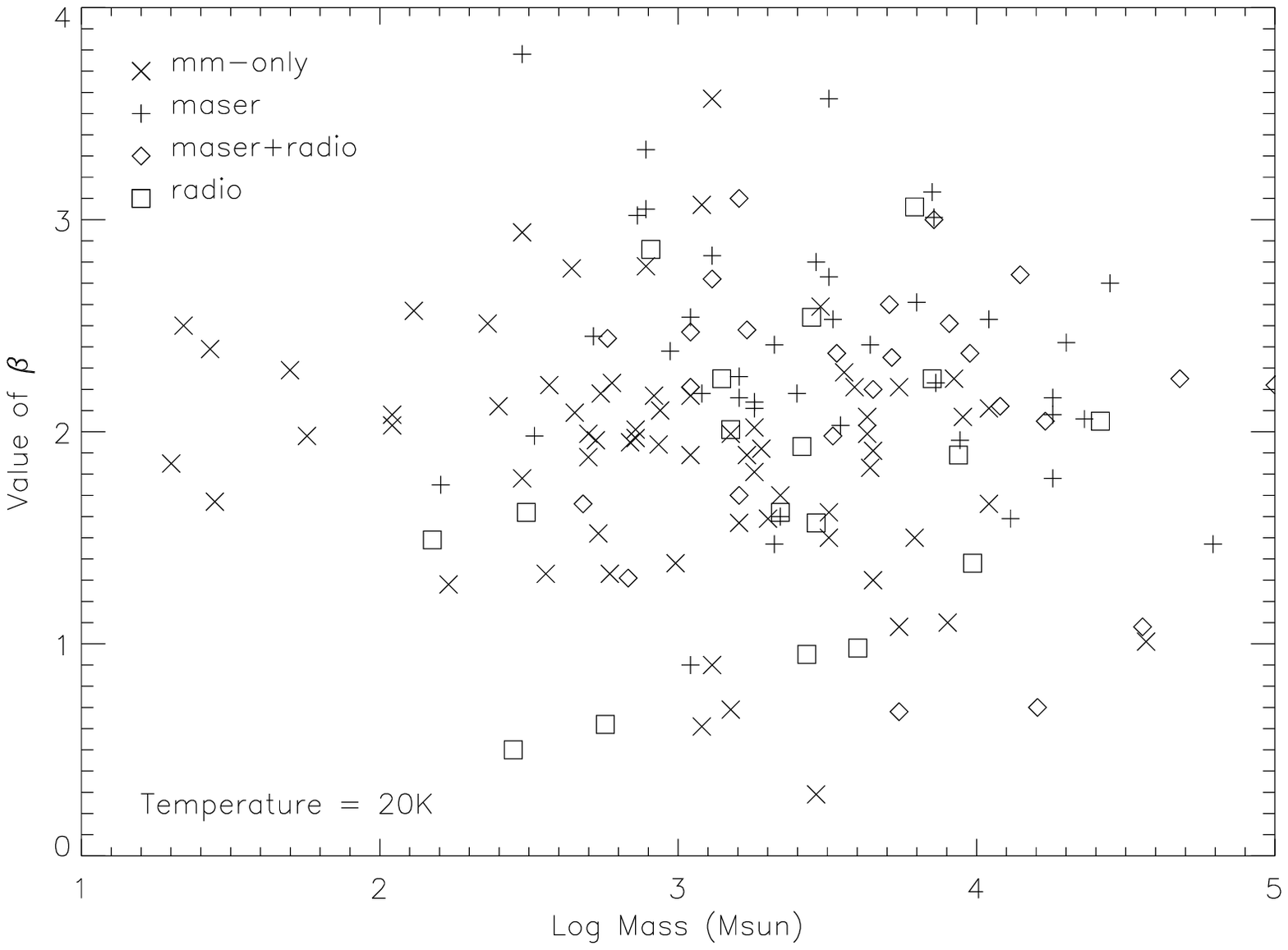}
	 \includegraphics[width=7.0cm, height=7.0cm]{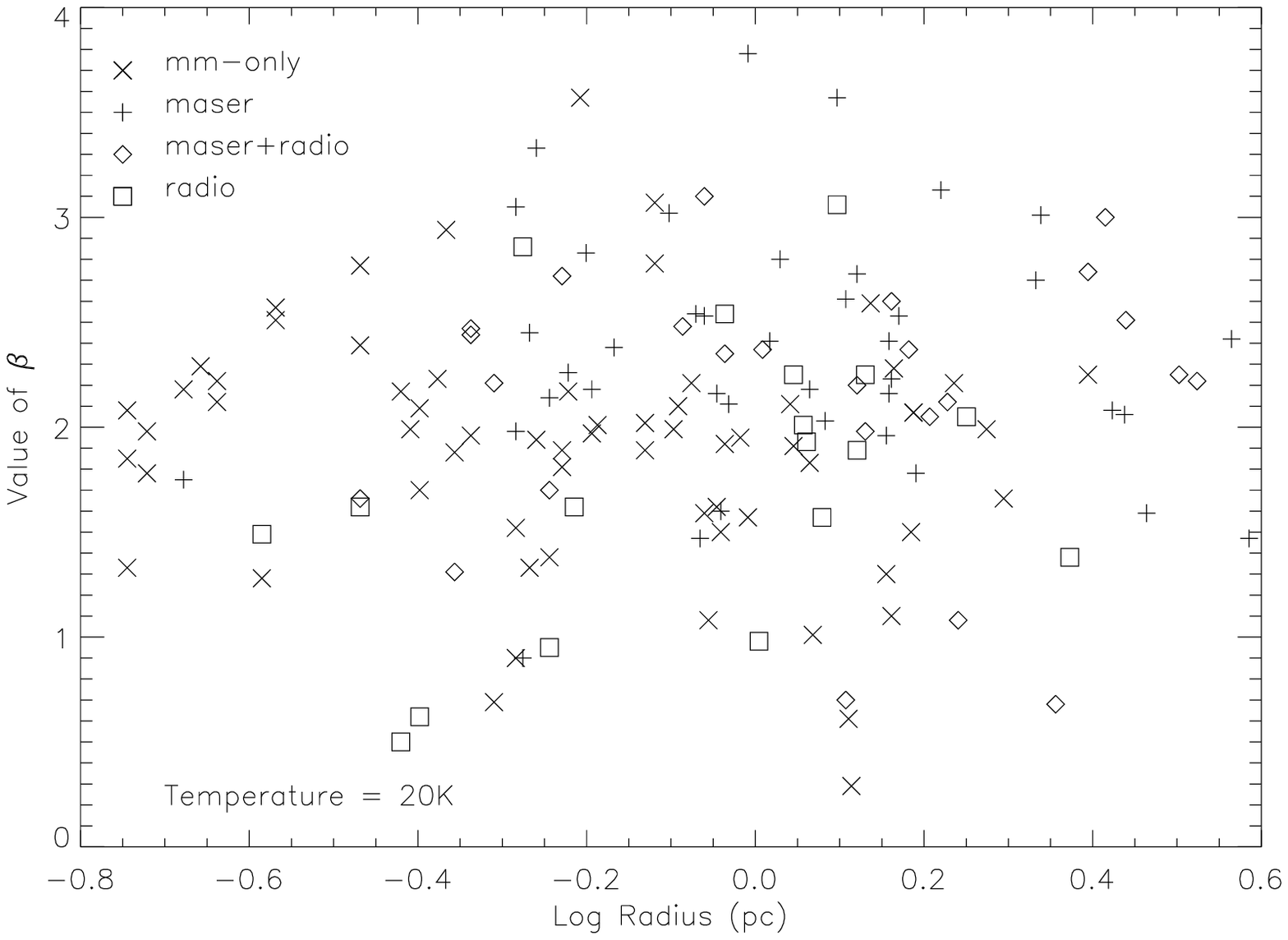}
  \end{minipage}
  \begin{minipage}{1.0\textwidth}
         \includegraphics[width=7.0cm, height=7.0cm]{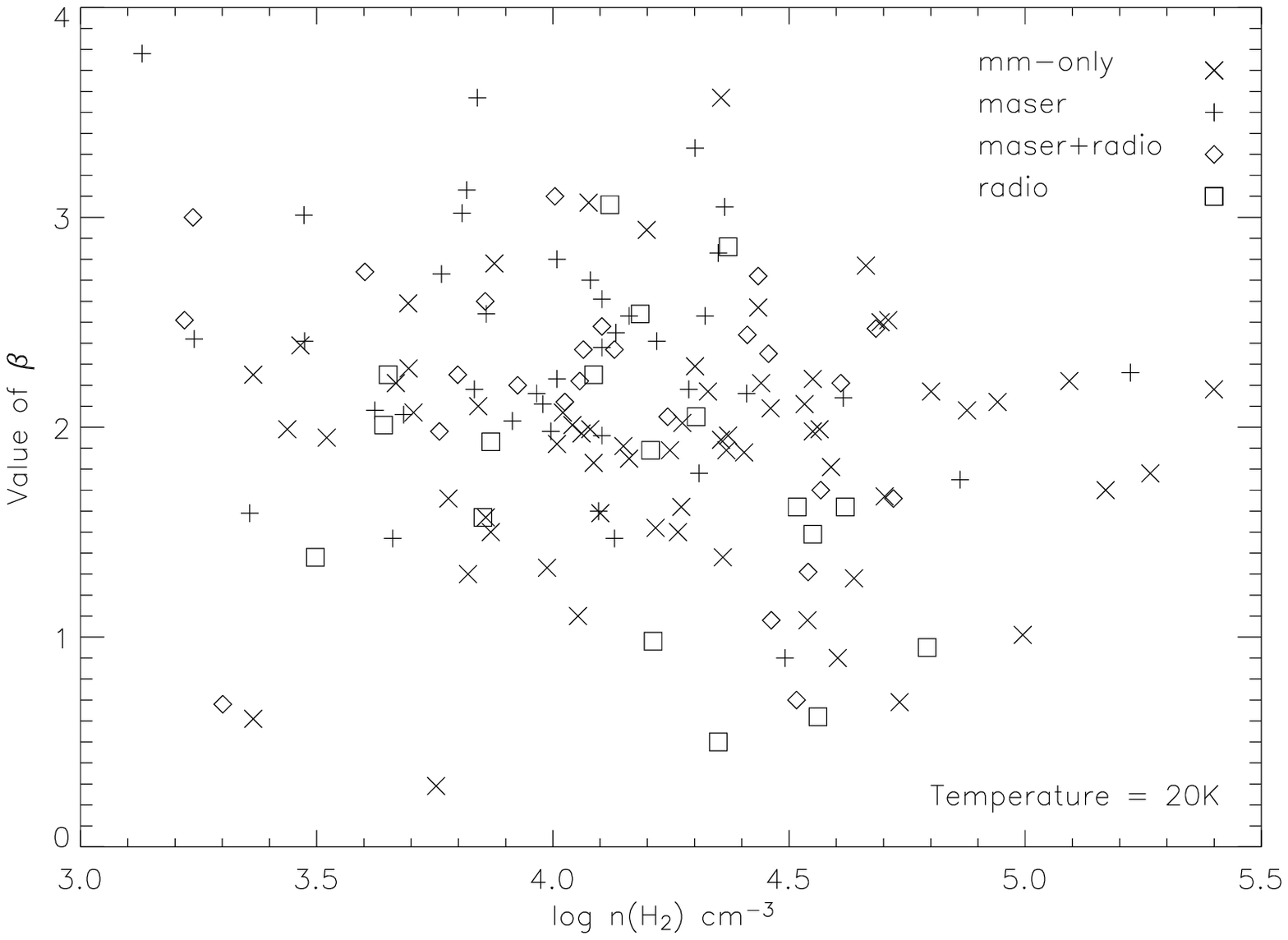}
	 \includegraphics[width=7.0cm, height=7.0cm]{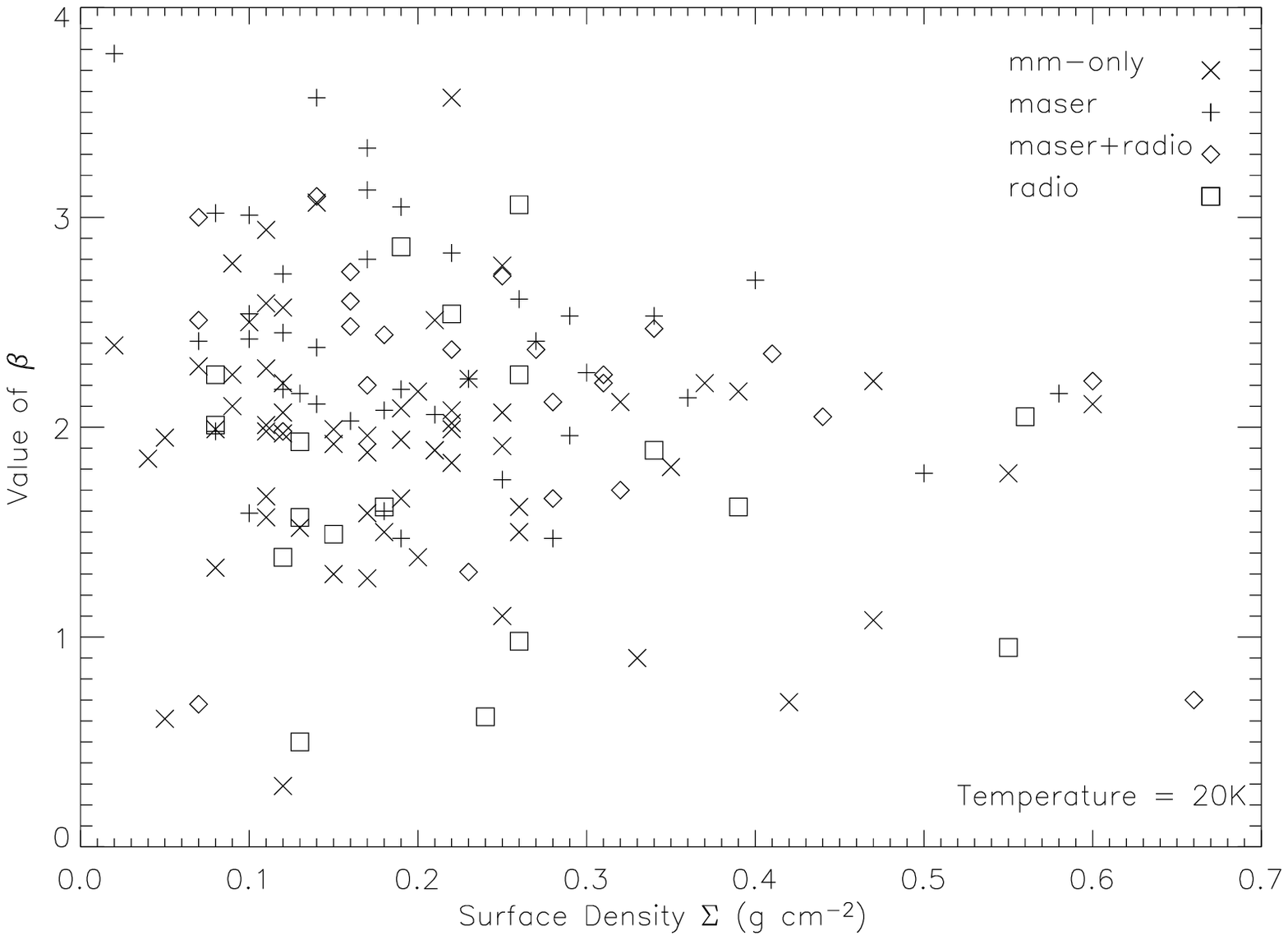}
  \end{minipage}
  \begin{minipage}{1.0\textwidth}
         \includegraphics[width=7.0cm, height=7.0cm]{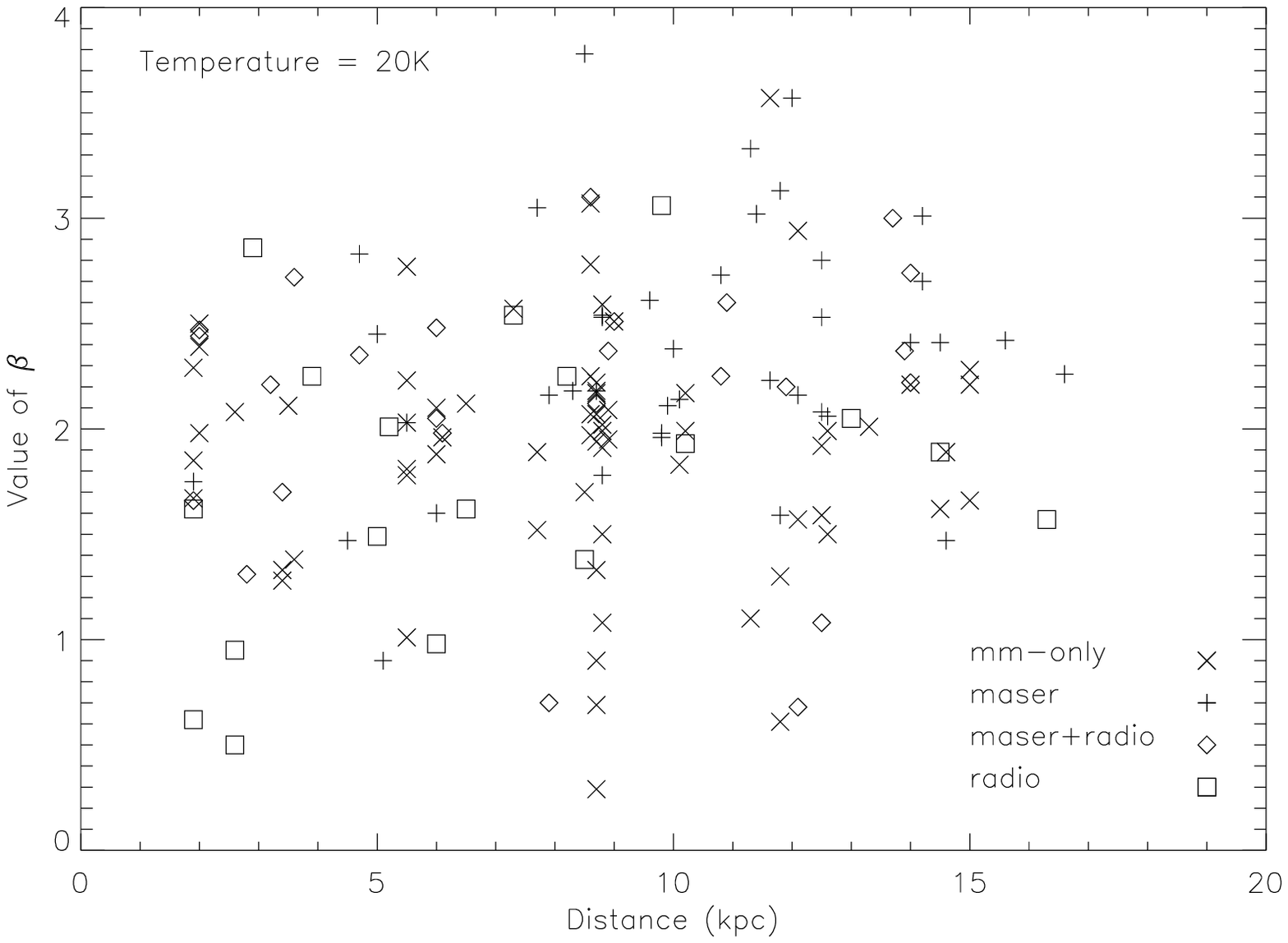}
  \end{minipage}
  \vspace{+0.1cm}

  \caption{Correlation plots of the dust grain emissivity index $\beta$ with mass, radius, H$_2$ number density ($n_{H_2}$), surface density ($\Sigma$) and the distance for an assumed temperature of 20K. The different classes of source are as depicted on the plots.  Top: (left) correlation plot of mass and $\beta$ of all sources (right) correlation plot of radius and $\beta$. Middle: (left) plot of $n_{H_2}$ and $\beta$ (right) plot of $\beta$ and $\Sigma$. Bottom: plot of $\beta$ with distance.}
\label{chap:scuba:correlations}
\end{figure*}

   We have also correlated the value of $\beta$ at 20\,K with the parameters of mass, radius, hydrogen number density  ($n_{H_{2}}$), surface density ($\Sigma$) and distance reported in Paper I, also determined with a temperature of 20\,K. The plots of each of these parameter correlations are presented in Figure \ref{chap:scuba:correlations}. It can be seen from these plots, that there is no obvious correlation between the value of $\beta$ and the other parameters tested. This is not unexpected as the physical state of the dust grains is probably not related to the source characteristics, at least for these early stages of star formation.

  From Figures \ref{chap:scuba:hist_temperature} and \ref{chap:scuba:hist_tracer} and Table \ref{chap:scuba:averages}, it can be seen that the dust grain emissivity index is typically of order 2. A value of $\beta$\,=\,2 is a reasonable value for $\beta$ for all source types, and for the sample in general, for assumed temperatures of 20\,-\,50\,K. This is not the case at 10\,K, however we do not expect many sources, if any, to have such low temperatures.

\subsection[]{The submillimetre sources of this survey}\label{chap:scuba:discussion:beta:submm}

\begin{table}
\begin{center}
\caption{The dust grain emissivity index $\beta$ for sources seen only at submillimetre wavelengths in this survey, for temperatures of 10\,-\,50\,K. \label{chap:scuba:submmbeta}}
\end{center}
\vspace{-0.5cm}
\begin{tabular} {@{}lccccc@{}}
\hline
            &  \multicolumn{5} {c}{Dust emissivity exponent $\beta$} \\         
Source Name& 10 & 20 & 30 & 40 & 50\\
\hline
G0.264+0.031 & 3.4 & 2.5 & 2.2 & 2.1 & 2.0	\\
G0.526+0.175 & 3.0 & 2.1 & 1.8 & 1.7 & 1.6	\\
G0.568-0.864 & 2.6 & 1.7 & 1.5 & 1.3 & 1.3	\\
G5.948-0.233 & 1.8 & 0.9 & 0.6 & 0.5 & 0.4	\\
G8.136+0.235$^\dagger$ & 2.1 & 1.2 & 1.0 & 0.8 & 0.8	\\
G10.615-0.336 & 3.1 & 2.1 & 1.9 & 1.8 & 1.7	\\
G12.927+0.485 & 2.1 & 1.1 & 0.9 & 0.8 & 0.7	\\
G11.935-0.159 & 2.4 & 1.4 & 1.2 & 1.1 & 1.0	\\
G12.205-0.125 & 1.6 & 0.7 & 0.4 & 0.3 & 0.2	\\
G12.709-0.223 & 2.5 & 1.6 & 1.3 & 1.2 & 1.1	\\
G16.879-2.180 & 3.1 & 2.1 & 1.9 & 1.8 & 1.7	\\
G16.883-2.188 & 3.6 & 2.7 & 2.4 & 2.3 & 2.2	\\
G16.887-2.191 & 2.3 & 1.4 & 1.2 & 1.0 & 1.0	\\
G23.688+0.158 & 2.9 & 2.0 & 1.7 & 1.6 & 1.5	\\
G23.695+0.165 & 3.6 & 2.7 & 2.5 & 2.3 & 2.3	\\
G23.695+0.157 & 3.4 & 2.5 & 2.2 & 2.1 & 2.0	\\
G23.694+0.145$^\dagger$ & 2.9 & 2.0 & 1.7 & 1.6 & 1.5	\\
G28.300+0.003 & 2.7 & 1.8 & 1.6 & 1.4 & 1.4	\\
G29.927-0.008 & 4.2 & 3.3 & 3.0 & 2.9 & 2.8	\\
G29.920-0.016$^\dagger$ & 3.0 & 2.0 & 1.8 & 1.7 & 1.6	\\
G29.915-0.023 & 3.2 & 2.2 & 2.0 & 1.9 & 1.8	\\
G29.918-0.054 & 1.8 & 0.9 & 0.7 & 0.5 & 0.5	\\
G29.932-0.050 & 2.5 & 1.6 & 1.3 & 1.2 & 1.1	\\
G29.924-0.054 & 2.3 & 1.4 & 1.1 & 1.0 & 0.9	\\
G29.943-0.049 & 2.6 & 1.7 & 1.4 & 1.3 & 1.3	\\
G29.919-0.065 & 2.2 & 1.3 & 1.1 & 0.9 & 0.9	\\
G30.884+0.153 & 1.4 & 0.5 & 0.2 & 0.1 & 0.0	\\
G30.881+0.141 & 2.0 & 1.0 & 0.8 & 0.7 & 0.6	\\
G30.903+0.148 & 2.0 & 1.1 & 0.8 & 0.7 & 0.6	\\
G30.712-0.074 & 3.3 & 2.4 & 2.2 & 2.0 & 2.0	\\
G31.238+0.640$^\star$ & 1.4 & 0.5 & 0.2 & 0.1 & 0.0	\\
G35.257+0.165 & 2.5 & 1.6 & 1.3 & 1.2 & 1.1	\\
G35.573+0.034 & 2.8 & 1.9 & 1.6 & 1.5 & 1.4	\\
G35.578+0.021 & 3.2 & 2.3 & 2.1 & 1.9 & 1.9	\\
G37.468-0.103 & 3.1 & 2.1 & 1.9 & 1.8 & 1.7	\\
G49.584-0.570 & 2.8 & 1.9 & 1.6 & 1.5 & 1.4	\\
G49.471-0.367 & 2.7 & 1.8 & 1.5 & 1.4 & 1.3	\\
G49.511-0.395 & 3.2 & 2.3 & 2.0 & 1.9 & 1.8	\\
G49.513-0.399 & 2.7 & 1.8 & 1.6 & 1.4 & 1.4	\\
G49.469-0.426 & 3.4 & 2.5 & 2.2 & 2.1 & 2.0	\\
\hline
mean & 2.7 & 1.8 & 1.5 & 1.4 & 1.3\\
\hline
\end{tabular}
\begin{flushleft}
$^\dagger$Denotes those submillimetre sources which appear as a single 850\microm\, source, but are resolved into two sources at the higher resolution 450\microm. In this instance, the 450\microm\, fluxes of the two sources have been added together.
$^\star$Denotes a source that has three peaks of emission at both 450 and 850\microm\, for which it is not possible to differentiate the flux for each of the components. These three sources have been treated as a single submillimetre source.
\end{flushleft}
\end{table}

   We have also determined the dust grain emissivity index $\beta$ for those submillimetre sources detected in this SCUBA survey, with flux densities measured at both 450 and 850\microm\, (40 sources) but devoid of 1.2-mm continuum emission. These sources are denoted by `submm' in column 4 of Table \ref{chap:scuba:scubafluxes}.  Only those sources with flux densities 5-$\sigma$ above the background noise are examined. We have employed a Levenberg-Marquardt least squares fit, as described in Section \ref{chap:scuba:beta_exp}, consistent with the $\beta$ analysis for the SIMBA sources mentioned above in Section \ref{chap:scuba:analysis} for a range of temperatures from 10\,-\,50\,K. The resultant value of $\beta$ for each temperature, together with a mean value of $\beta$ at each of the temperatures, is presented in Table \ref{chap:scuba:submmbeta}. The standard deviation of $\beta$ at each temperature is 0.6.

   It can be seen from Table \ref{chap:scuba:submmbeta} that the mean value of $\beta$ decreases with an increase in assumed temperature when fitting, which is consistent with the results for the SIMBA sources in Section \ref{chap:scuba:analysis}. The mean value of $\beta$ for these submillimetre sources is slightly lower than that determined when millimetre continuum data was also available, although the difference is within the standard deviation of the sample (see Table \ref{chap:scuba:averages}).

   However, the emissivity index would also be consistent with the value $\beta$\,=\,2 if the temperature of these submillimetre only sources was somewhat lower, in the range 10\,-\,20\,K, than those sources with millimetre data as well. Such colder cores would also have lower fluxes, so such a result is not unexpected.

\subsection[]{The Dust Grain Emissivity Index $\beta$}\label{chap:scuba:discuss:beta}

   The dust grain emissivity index is crucial in determining many source parameters, including the dust temperature, as well as providing information about the grain structure and the dielectric characteristics of the grains \citep{schwartz82, hildebrand83}. Knowledge of the dust grain emissivity index ($\beta$) improves estimates of the dust opacity, which in turn bears heavily on the dust mass of the star forming cloud from observations of the radiation of the dust grains. If the dust properties are constrained, then the mass of the star forming cloud can be determined more precisely. Determination of $\beta$ will therefore facilitate future spectral energy distribution analysis (to be presented in a forthcoming paper).

   Various estimates of the dust grain emissivity index $\beta$ from both laboratory and theoretical models have been proposed. These models include estimates of what $\beta$ should be for different chemical compositions, which are thought to represent different sized interstellar grains at different temperatures. However, there is little consensus between these studies.

   \citet{draine84} used available laboratory data to measure optical constants for a mixture of naked graphite and silicate grains. They found that $\beta$\,$\sim$\,2 for 40\,$\le\,\lambda\,\le$\,1000\,$\mu$m. A dust grain emissivity index equivalent to a value of 2 has also been supported in the calculations of \citet{knacke73, mathis89, kruegel94}. However, the largest grains ($>$\,30\,$\mu$m) in the sample of \citet{kruegel94} did not follow the trend of $\beta$\,=\,2 as per the rest of the sample.  \citet{miyake93} determined values of $\beta$ in excess of 2 for large grain sizes. \citet{mathis89} found $\beta$\,=\,1.5 for composite grains, whilst \citet{wright87} calculated 0.6\, $\le\,\beta\,\le$\, 1.4 for fractal grains. Grain models by \citet{aannestad75} predict $\beta$ up to 3.5 for olivine, fused quartz, and lunar rock core grains covered with ice mantles. \citet{koike95} found values of the dust emissivity exponent up to a value of 2.8. \citet{mennella95} conclude that more evolved cores have a higher emissivity exponent ($\beta$). 

   A dust emissivity exponent equivalent to 2, has been adopted in the models of \citet{hildebrand83} and \citet{gordon95}, and is generally assumed when calculating the dust mass for an unknown $\beta$ \citep[cf.][]{dunne00, james02, kramer03}.

  Observationally, there has been an abundance of data attempting to determine and explain $\beta$.  Observationally determined values of the dust emissivity exponent also vary quite notably, with most findings between 1 and 2. Previous observational studies of $\beta$ have tended to focus on a particular region or source.

 \citet{lismenten98} found $\beta$\,=\,2 in their observations of the giant molecular cloud GMC\,0.25+0.11, while \citet{gordon88} examined warm cloud complexes associated with \hii\, regions and found an emissivity exponent equivalent to 2. Further support for $\beta$ = 1.5 - 2.0 include observations by \citet{gezari73, soifer75, schwartz82, wright92, masi95, minier05}.

  Support for more extreme values of $\beta$ than 2 include, \citet*{kuan96} who found that Sgr B2 had an emissivity index $\beta$ of 3.7 (at $\sim$~3~mm), which they interpreted as strong evidence for ice-coated core-mantle grains. \citet{dowell99} found that $\beta$ near the Galactic Centre was equivalent to 2, but was as high as 2.5 in parts of the dust ridge. \citet{gordon90} derived an emissivity exponent of 2.2 for the Cepheus A and B molecular clouds. \citet{p-p00} found a value of 2.4 for $\beta$ in their submillimetre study of the Galactic Centre region, while \citet{burton04} determined an emissivity index of 3.6 for an isolated high-mass young stellar object near the Galactic Centre.

   Lower values of $\beta$ include the observations of \citet{oldman94}, who examined the W3 star forming cloud, finding an emissivity index $\beta$ = 1.4 in the vicinity of sources, but more generally, the cloud was determined to have a $\beta$ $\sim$~1.0. More recently, \citet{williams04} found an average emissivity index of  $\beta$ = 0.9  for their sample of 68 high-mass protostellar objects. Observations of young and evolved stars result in  low values of $\beta$ $\sim$ 0.2 - 1.4 \citep*[e.g][]{weintraub89, knapp93}. There has also been extensive work on determining the dust grain emissivity for nearby galaxies \citep[e.g.][]{chini89, chini93, bianchi98, dunne00, dunne01}.

   \citet{goldsmith97} found variation of the dust grain emissivity exponent relative to the density gradient of the source for two of the three sources that they examined. M17 displayed a small variation of $\beta$ across the core, while Cepheus A also displayed a small reduction of this exponent between the core of the source and its edge. These results indicate that $\beta$ is smaller as one goes to regions of higher density. It would be useful to examine variation in the value of $\beta$ across the cores in our sample, however, the spatial resolution of the observations utilised in this analysis prevent this. With higher spatial resolution (e.g. ALMA), it would be possible and useful to examine a variation of $\beta$ with respect to the density gradient of the cores in our sample.

   There have been few large scale observational studies of the dust grain emissivity index $\beta$ as per this one. Instead, many prefer to assume a value of $\beta$ when it is unknown. Our results show that $\beta$ is $\sim$\, 2, which is consistent with values reported throughout the literature, as discussed in the preceding paragraphs.

   The value of $\beta$ provides an indication of the type of grains present in the central star forming complex, e.g. $\beta$\,=\,2 as modelled by \citet{draine84} for a mixture of graphite and silicate grains. Our work extends considerably the sample used to determine the value of this index.

\section{Conclusions}

   We have identified 212 sources in this survey, half of which are known SIMBA millimetre continuum sources. The remaining 106 sources are submillimetre cores, 50 per cent of which resolve a known SIMBA source into multiple submillimetre components, whilst the other 53 submillimetre sources are not associated with SIMBA millimetre continuum emission. Additionally, we have identified two further mm-only sources in the SIMBA images, thereby increasing the SIMBA source list to 405, and the number of mm-only cores to 255.

   We have concatenated the results from four (sub)millimetre surveys \citep[][as well as this work]{walsh03, hill05, thompson06}, with the Galactic Plane map of \citet{p-p00} in order to examine the dust grain emissivity exponent $\beta$ for SIMBA sources reported in Paper I. We have applied a Levenberg-Marquardt least squares fit to a total of 154 sources of the SIMBA source list in order to observationally determine the value of $\beta$, for an assumed temperature. We find that the dust grain emissivity exponent $\beta$ is typically of order 2, which is characteristic of graphitic grains, and with values reported throughout the literature. Determination of this emissivity exponent will facilitate future spectral energy distribution analysis for the SIMBA source list reported in Paper I, by constraining the dust properties and hence the dust mass.
 
   We have correlated the parameters of mass, radius, H$_2$ number density ($n_{H_2}$), surface density ($\Sigma$) and distance, with respect to the dust grain emissivity index $\beta$. There is no obvious correlation of $\beta$ with these parameters, as is expected.

\section*{Acknowledgements}

   The authors would like to thank the staff at the Joint Astronomy Centre (JAC) for their helpful comments and recommendations, in particular Harold Butner and Gerald Moriarty-Schieven. We thank Jim Hoge and Jan Waterlout for their observational support and Remo Tilanus for help with ORACDR. We are grateful to Alex Pope of the University of British Columbia for help with SURF. We also thank Melvin Hoare for discussion on the dust emissivity exponent. We thank an anonymous referee for the positive feedback and useful recommendations.

   The authors extend thanks to the Australian Research Council (ARC), and acknowledge the AMRF program of the Australian Nuclear Science and Technology Organisation (ANSTO) for travel support to the JCMT in May 2005. T.H is also grateful to the UNSW Faculty Research Grant Program for funding support and the Australia Telescope National Facility (ATNF), a division of the CSIRO for their support.
   
    The data were reduced using the standard SCUBA User Reduction Facility (SURF) package developed by T. Jenness and J. F. Lightfoot. This work has made use of the Starlink image analysis packages KAPPA and GAIA, as well as the image production toolkit KARMA, in conjunction with the IDL computing package.
   
\bibliographystyle{mn2e}
\bibliography{aa,references}
\expandafter\ifx\csname natexlab\endcsname\relax\def\natexlab#1{#1}\fi

\appendix
\section[]{Presentation of the Images}

   Maps of the 450 and 850\microm\, submillimetre continuum emission detected with the SCUBA instrument on the JCMT are presented here. Coordinates of the images are in J2000 epoch. The images are oriented with the right ascension on the horizontal and the declination on the vertical. The methanol maser is depicted as a `plus' symbol in these images, while the radio continuum (\uchii) source is denoted by a `box' symbol, which is consistent with Figure \ref{chap:scuba:scubasample}. Class MR objects (with both a methanol maser and an \uchii\, region) will house both a `plus' and a `box' symbol. The source at the centre of the image is the mm-only source of the SIMBA survey, which was targeted here. Source names are depicted in the top-left corner of the images, whilst the beam size is included in the lower left corner. Contour levels are as described in the text (see Section \ref{chap:scuba:sampleimages}). `Holes' in the images are indicative of bad/noisy bolometers during the jigglemap scans, which have been removed in the data reduction process. 

\bsp

\label{lastpage}

\end{document}